\documentclass[]{article}

\title{The Scattering Map on Collapsing Charged Spherically Symmetric Spacetimes}
\author{Fred Alford}
\date{Department of Mathematics,\\
	Imperial College London\\
	[2ex]\today}
\newcommand{\p}{\partial}
\newcommand{\R}{\mathbb{R}}
\usepackage{amsmath,amssymb,amsthm}
\usepackage{geometry}
\usepackage{slashed}
\usepackage{array}
\usepackage{tikz}
\usepackage{wrapfig}
\usepackage{graphicx}
\usepackage{cutwin}
\usepackage[numbers]{natbib}
\usepackage{bbm}
\usepackage{floatrow}
\usetikzlibrary{decorations.pathmorphing}
\newtheorem{Theorem}{Theorem}[section]
\newtheorem{Lemma}[Theorem]{Lemma}
\newtheorem{thm}{Theorem}

\newtheorem{Proposition}[Theorem]{Proposition}
\newtheorem{Remark}[Theorem]{Remark}

\usetikzlibrary{decorations.pathmorphing}
\tikzset{snake it/.style={decorate, decoration=snake}}
\usepackage{xpatch}
\makeatletter
\xpatchcmd{\@thm}{\thm@headpunct{.}}{\thm@headpunct{}}{}{}
\makeatother
\numberwithin{equation}{section}
\begin{document}

\maketitle

\begin{abstract}
	In this paper we generalise our previous results \cite{Mine} concerning scattering on the exterior of collapsing dust clouds to the charged case, including in particular the extremal case. We analyse the energy boundedness of solutions $\phi$ to the wave equation on the exterior of collapsing spherically symmetric charged matter clouds. We then proceed to define the scattering map on this spacetime, and look at the implications of our boundedness results on this map.
	
	More specifically, we first construct a class of spherically symmetric charged collapsing matter cloud exteriors, and then consider solutions to the wave equation with Dirichlet (reflective) boundary conditions on the surface of these clouds. We then show that the energy of $\phi$ remains uniformly bounded going forwards or backwards in time, and that the scattering map is bounded going forwards but not backwards. Therefore, the scattering map is not surjective onto the space of finite energy on $\mathcal{I}^+\cup\mathcal{H}^+$. Thus there does not exist a backwards scattering map from finite energy radiation fields on $\mathcal{I}^+\cup\mathcal{H}^+$ to finite energy radiation fields on $\mathcal{I}^-$ for these models.
	
	These results will be used to give a treatment of Hawking radiation in a companion paper \cite{Mine3}.
\end{abstract}

\section{Overview}

In \cite{Mine}, we initiated the study of the classical scattering of waves on fully dynamical collapsing spacetimes, specifically the Oppenheimer--Snyder model. This plays a significant role in the mathematical study of Hawking radiation \cite{hawking1975}. Because extremal black holes play a distinguished role within the study of Hawking radiation, it is important to include these in our models. To this end, this paper looks at the scattering map for massless scalar waves on a class of spherically symmetric, charged, collapsing spacetime models, which can be viewed as a generalisation of the Oppenheimer--Snyder model. This class includes models which collapse to form both sub-extremal and extremal Reissner--Nordstr\"om black holes \cite{Reissner}. This will allow us to study Hawking radiation in a mathematical context in a companion paper, \cite{Mine3}.

For the convenience of the reader, we closely follow the structure of our previous paper \cite{Mine}. We hope this will make similarities and differences between the charged and uncharged case more clear.

In this paper we will be studying the energy boundedness of solutions to the linear wave equation
\begin{equation}\label{eq:wave}
\Box_g\phi:=\frac{1}{\sqrt{-g}}\p_a\left(\sqrt{-g}g^{ab}\p_b\phi\right)=0
\end{equation}
on collapsing spherically symmetric spacetimes, i.e.~solutions to the Einstein--Maxwell equations outside an evolving sphere. This sphere is given by $\{(t^*, r_b(t^*),\theta,\varphi)\}$ in the coordinates below with some restrictions placed on the function $r_b$. As this exterior is an asymptotically flat solution of the Einstein--Maxwell equations, it has a Reissner--Nordstr\"om metric, given by
\begin{align}
g=-\left(1-\frac{2M}{r}+\frac{q^2 M^2}{r^2}\right){dt^*}^2+2\left(\frac{2M}{r}-\frac{q^2 M^2}{r^2}\right)&dt^*dr+\left(1+\frac{2M}{r}-\frac{q^2 M^2}{r^2}\right)dr^2+r^2g_{S^2}\\\nonumber
t^*\in\R\qquad& r\in[\tilde{r}_b(t^*),\infty),
\end{align}
where $g_{S^2}$ is the metric on the unit $2$-sphere, $\tilde{r}_b=\max\{r_b,r_+\}$, and $r=r_+$ is the horizon of the underlying Reissner--Nordstr\"om metric as given by \eqref{eq:r+}. The parameter $M$ is positive, and the parameter $q$ takes values in the range $[-1,1]$, with $q^2=1$ corresponding to the extremal case.

These collapsing matter cloud models will include the Oppenheimer--Snyder Model \cite{S-O}, and we will refer to these more general models as Reissner--Nordstr\"om Oppenheimer--Snyder (RNOS) models. These will include both extremal and sub-extremal cases.

We will be imposing Dirichlet (i.e.~reflective) conditions on the boundary of the matter cloud, i.e.~$\phi=0$ on $r=r_b$ (in a trace sense), and then proceed to define a scattering theory for these spacetimes. In \cite{Mine}, we considered both the permeating and the reflective cases. However, here we will not attempt to consider the interior of our matter cloud as this will depend entirely on one's choice of matter model.

The main theorems of this paper are informally stated below:

\begin{thm}[Uniform Non-degenerate Energy Boundedness]\label{Thm:VagueBound}
	For all RNOS models, including the extremal case $\vert q\vert=1$, we define $\mathcal{F}_{[t^*_0,t^*_1]}$, $t^*_0\leq t^*_1$, to be the map taking solutions of $\eqref{eq:wave}$ on $\Sigma_{t^*_0}$ forward to the same solution evaluated on $\Sigma_{t^*_1}$. Then $\mathcal{F}_{[t^*_0,t^*_1]}$ is uniformly bounded with respect to the non-degenerate energy of $\phi$. Furthermore, for $t^*_1\leq t^*_c$, the inverse of $\mathcal{F}_{[t^*_0,t^*_1]}$ is also uniformly bounded with respect to the non-degenerate energy.
	
	This Theorem is stated more precisely across Theorems \ref{Thm:T^*<1}, \ref{Thm:T^*>1}, \ref{Thm:BackwardsBound}.
\end{thm}

\begin{wrapfigure}{r}{5cm}\label{Fig:Penrose}
	\begin{tikzpicture}[scale =1.2]
	\node (I)    at ( 0,0) {};
	
	\path 
	(I) +(90:2)  coordinate[label=90:$i^+$]  (Itop)
	+(-90:2) coordinate (Imid)
	+(0:2)   coordinate[label=0:$i^0$] (Iright)
	+(-1,1) coordinate (Ileft)
	+(-0.6,1.4) coordinate[label=0:\tiny ($t^*_c$\text{, }$r_+$)] (BHH)
	+(-1,-3) coordinate[label=0:$i^-$] (Ibot)
	;
	\draw (Ileft) -- 
	node[midway, above left]    {$\mathcal{H}^+$}
	(Itop) --
	node[midway, above, sloped] {$\mathcal{I}^+$}
	(Iright) -- 
	node[midway, below, sloped] {$\mathcal{I}^-$}
	(Ibot) --
	node[midway, above, sloped]    {\small }    
	(Ileft) -- cycle;
	\draw[fill=gray!80] (Ibot) to[out=60, in=-60]
	node[midway, below, sloped] {\tiny $r=r_b$} (BHH)--(Ileft)--cycle;
	\draw (Iright) to[out=170, in=0] node[midway, above, sloped] {\tiny $\Sigma_{t^*}$} (-0.23,0.5);
	\end{tikzpicture}
	\caption{Penrose Diagram of RNOS Model, with spacelike hyper surface $\Sigma_{t^*}$.}\label{fig:PenRef}
\end{wrapfigure}
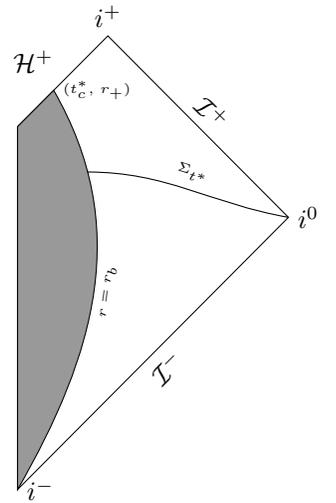

The hypersurface $\Sigma_{t^*}$ is shown in Figure \ref{Fig:Penrose}, as is the sphere $(t^*_c,r_+)$. We define $t^*_c$ as the $t^*$ coordinate for which $r_b(t^*_c)=r_+$.

Here ``non-degenerate energy'' signifies energy flux through the surface with respect to an \textit{everywhere timelike} vector field (including at the boundary of the matter cloud and the horizon), which coincides with the timelike Killing field in a neighbourhood of null infinity. Non-degenerate energy through the surface $\Sigma_{t^*}$ controls the $L^2$ norm of $\phi$'s derivatives.

We next turn to defining a radiation field on future and past null infinities $\mathcal{I}^\pm$ and at the future horizon $\mathcal{H}^+$. Note that previous works on Reissner--Nordstr\"om black holes, see \cite{Moschidis} for example, already give us the existence of the future radiation field at $\mathcal{I}^+$ and $\mathcal{H}^+$, so we only need to concern ourselves with the past radiation field on $\mathcal{I}^-$:

\begin{thm}[Existence and Non-degenerate Energy Boundedness of the Past Radiation Field]\label{eq:VagueExist}
	For all RNOS models, including the extremal case $\vert q\vert=1$, we define the map $\mathcal{F}^-$, which takes solutions of \eqref{eq:wave} from $\Sigma_{t^*_c}$ to their past radiation field on $\mathcal{I}^-$. Then $\mathcal{F}^-$ exists and is bounded with respect to the non-degenerate energy. The inverse of $\mathcal{F}^-$, denoted $\mathcal{F}^+$, is also bounded with respect to the non-degenerate energy. Finally, $\mathcal{F}^-$ is a bijection between these finite energy spaces.
	
	This Theorem is stated more precisely in Theorem \ref{Thm:Fbijection}.
\end{thm}

We finally define the map $\mathcal{G}^+$, which takes solutions on $\Sigma_{t^*_c}$ to their radiation field on $\mathcal{H}^+\cup\mathcal{I}^+$. We make use of previous results from \cite{BHL,ERNScat}, from which we know $\mathcal{G}^+$ is bounded, and that $\mathcal{G}^-:=(\mathcal{G}^+)^{-1}$ (only defined on the image of $\mathcal{G}^+$) is unbounded. This leads us to the final theorem on the scattering map:

\begin{thm}[Boundedness and Non-surjectivity of the Scattering Map]
	For all RNOS models, including the extremal case $\vert q\vert=1$, we define the scattering map 
	\begin{equation}
	\mathcal{S}^+:=\mathcal{G}^+\circ\mathcal{F}^+.
	\end{equation} 
	This map takes the radiation field of a solution to \eqref{eq:wave} on $\mathcal{I}^-$ to that solution's radiation fields on $\mathcal{H}^+\cup\mathcal{I}^+$. Then $\mathcal{S}^+$ is bounded with respect to the non-degenerate energy ($L^2$ norms of $\p_v(r\phi)$ on $\mathcal{I}^-$ and $\mathcal{H}^+$ and $\p_u(r\phi)$ on $\mathcal{I}^+$). However, if we define the inverse of $\mathcal{S}^+$, denoted by $\mathcal{S}^-$ (only defined on the image of $\mathcal{S}^+$), this is not bounded with respect to the non-degenerate energy. Therefore there does not exist a backwards scattering map going from finite energy spaces on $\mathcal{H}^+\cup\mathcal{I}^+$ back to finite energy spaces on $\mathcal{I}^-$.
	
	This Theorem is stated more precisely as Theorem \ref{Thm:ScatBound}.
\end{thm}

The non-invertibility of $\mathcal{S}^+$ arises from the non-invertibility of $\mathcal{G}^+$, exactly as in the uncharged case \cite{Mine}. It is the existence of the map $\mathcal{F}^+$ mapping into the space of non-degenerate energy that extends this non-invertibility to data on $\mathcal{I}^-$, and thus causes the collapsing case to differ from the Reissner--Nordstr\"om case.

\begin{Remark}
It is particularly surprising to note the non-surjectivity of $\mathcal{S}^+$ includes the extremal case. This occurs despite the significant differences in the properties of the black hole horizon, most notably the absence of the usual red-shift effect \cite{RedShift} in the extremal case.
\end{Remark}

It remains an open problem to precisely characterise the image of the scattering map $\mathcal{S}^+$, even in the uncharged case.
\subsection{Acknowledgements}
We would like to thanks Mihalis Dafermos for many insightful comments, and for proof reading the manuscript. We would also like to thank Owain Salter Fitz-Gibbon for many insightful discussions. Last but by no means least, we would like to thank Claude Warnick and Bernard Kay for their comments and suggestions.

This work was part funded by EPSRC DTP, 
$[1936235]$. This work was supported by the Additional Funding Programme for
Mathematical Sciences, delivered by EPSRC (EP/V521917/1) and the
Heilbronn Institute for Mathematical Research.

\section{Previous Work}
There have been several previous works studying black hole scattering on collapsing spacetimes, see \cite{Bachelot,DiracScatter,BachelotHawking,Melnyk2004}. However, scattering in the collapsing charged case does not appear to have been considered previously. For a longer discussion of the uncharged case, we refer the reader to our previous work \cite{Mine} and references therein. There have, however, been other works considering the underlying models of charged collapse, and there have been other works considering scattering on Reissner--Nordstr\"om backgrounds. 

Several papers look at models of spherical collapse to Reissner--Nordstr\"om, such as \cite{RNCollapseModel,ChargedDustSolutionofRuban,GravCollapsewithCharge}. Most papers considering collapsing models focus on the interior of the collapsing star. This paper, however, will not focus on the specifics of interior models such as these, unlike \cite{Mine}. We note that there are many such models, which entirely depend on what equation of state is chosen for the interior of the matter cloud. 

Generally, study of the scattering map in the exterior sub-extremal Reissner--Nordstr\"om spacetime is paired together with that of Schwarzschild, as it has similar behaviour (see \cite{BHL}). The extremal case has been studied in detail separately, see \cite{ERNScat}, as behaviour in this case differs from the sub-extremal case. Scattering in the black hole interior has also been studied independently, \cite{InteriorScat}. The exterior of the RNOS Models (see Section \ref{Sec:RNOS}) is a spherically symmetric, vacuum solution to the Einstein--Maxwell equations, and thus has the Reissner--Nordstr\"om metric, by uniqueness (see \cite{RNUniqueness}, for example). However, this paper will not be discussing the scattering map on Reissner--Nordstr\"om much beyond this, and instead will quote results from \cite{BHL} (in the sub-extremal case) and \cite{ERNScat} (in the extremal case). We refer the reader to these for a more complete discussion of scattering in Reissner--Nordstr\"om spacetimes.

\section{RNOS Models}\label{Sec:RNOS}
In this section we look at our background models of spherically symmetric charged matter cloud collapse. In section \ref{Manifold Derivation}, we derive the metric of our spacetime in the exterior of our collapsing matter cloud. If the reader is not interested in this derivation, they may skip straight to section \ref{Manifold Definition}, where the background manifold is defined, with some interesting and/or useful properties stated.

\subsection{Example of a Physical RNOS model}\label{Manifold Derivation}
In this section, we derive a physical example of an RNOS model under the following assumptions: We assume our manifold is a spherically symmetric solution of the Einstein--Maxwell equations

\begin{align}\label{eq:EE}
R_{\mu\nu}-\frac{1}{2}Rg_{\mu\nu}&=8\pi \boldsymbol{T}_{\mu\nu}\\
\boldsymbol{T}_{\mu\nu}&=\frac{1}{4\pi}\left(F_{\mu\alpha}{F_\nu}^{\alpha}-\frac{1}{4}F^{\alpha\beta}F_{\alpha\beta}g_{\mu\nu}\right)\\
\nabla_\mu F^{\mu\nu}&=0\\
\label{eq:Max}
\nabla_\mu F_{\nu\rho}+\nabla_\nu& F_{\rho\mu}+\nabla_\rho F_{\mu\nu}=0,
\end{align}
with coordinates $t^*\in(-\infty,\infty)$, $(\theta,\varphi)\in S^2$, $r\in[\tilde{r}_b(t^*),\infty)$. We define
\begin{align}
\tilde{r}_b(t^*)&:=\begin{cases} r_b(t^*) & t^*\leq t^*_c\\
	r_+ &t^*>t^*_c\end{cases}\\\label{eq:r+}
r_+&:=M(1+\sqrt{1-q^2}).
\end{align} 
\begin{wrapfigure}{r}{7cm}
	\begin{tikzpicture}[scale =1.4]
		\node (I)    at ( 0,0) {};
		
		\path 
		(I) +(0,2)  coordinate[label=90:$i^+$]  (Itop)
		+(0,-2) coordinate (Imid)
		+(2,0)   coordinate[label=0:$i^0$] (Iright)
		+(-2,0) coordinate[label=180 :$\mathcal{B}$] (Ileft)
		+(0,-2) coordinate[label=0:$i^-$] (Ibot)
		;
		\draw (Ileft) -- 
		node[midway, above, sloped]    {$\mathcal{H}^+$, $r=r_+$}
		(Itop) --
		node[midway, above, sloped] {$\mathcal{I}^+$}
		(Iright) -- node[midway, below, sloped]    {$\mathcal{I}^-$}
		(Ibot) --
		node[midway, below, sloped]    {$\mathcal{H}^-$, $r=r_+$} 
		(Ileft) -- cycle;
		\draw[fill=gray!80] (Ibot) to[out=70, in=-60]
		node[midway, above, sloped] {\tiny $r=r_b$} (-0.5,1.5) --(Ileft)--cycle;
	\end{tikzpicture}
	\caption{Penrose diagram of pure Reissner--Nordstr\"om spacetimes}\label{fig:RNPenrose}
\end{wrapfigure}
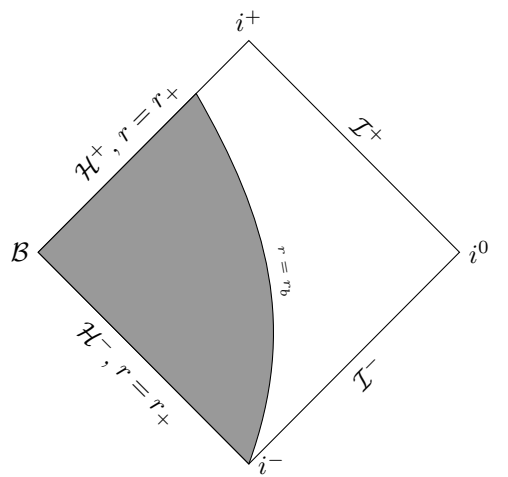

Here $r=r_b(t^*)$ is a hypersurface generated by a family of timelike, ingoing radial curves such that, for any fixed $\theta, \varphi$, the curve $\{t^*, r_b(t^*), \theta, \varphi\}$ describes the motion of a particle moving only under the electromagnetic force with fixed charge to mass ratio. the value of $t^*$ for which $r_b(t^*)=r_+$ is labelled $t^*_c$. That is, we assume that the surface of the cloud is itself massive and charged, with uniform charge to mass ratio across the surface. For our results we will actually only require certain bounds on $r_b$ and $\dot{r}_b$, but here we will consider one possible behaviour of $r_b$ in full. We also note that we are looking solely at the exterior of the black hole. Thus, we will not be considering the region $r_b(t^*)<r_+$. We will instead use $\tilde{r}_b(t^*)$ as the boundary of our manifold. The topology of our manifold $\{t^*,r\geq \tilde{r}_b(t^*),\theta,\varphi\}$ is that of the exterior of a cylinder in $3+1D$ Lorentzian space. As this is simply connected, equation \eqref{eq:Max} means we can choose an $A$ such that
\begin{equation}\label{eq:A Definition}
F=dA.
\end{equation}

Given an asymptotically flat, spherically symmetric solution of the Einstein--Maxwell equations, we know that our solution is a subset of a Reissner--Nordstr\"om spacetime (see for example \cite{RNUniqueness}). This gives the first two parameters of our spacetime; $M$, the mass of the Reissner--Nordstr\"om black hole spacetime our manifold is a subset of, and $q=Q/M$, the charge density of our underlying Reissner--Nordstr\"om spacetime. We will assume $q$ has modulus less than or equal to $1$, as otherwise our matter cloud will either not collapse, or will form a naked singularity rather than a black hole.

Exterior Reissner--Nordstr\"om spacetime has global coordinates:
\begin{equation}
	\mathcal{M}_{RN}=\R \times [r_+,\infty)\times S^2
\end{equation}
\begin{equation}
g=-\left(1-\frac{2M}{r}+\frac{q^2 M^2}{r^2}\right){dt^*}^2+2\left(\frac{2M}{r}-\frac{q^2 M^2}{r^2}\right)dt^*dr+\left(1+\frac{2M}{r}-\frac{q^2 M^2}{r^2}\right)dr^2+r^2g_{S^2}
\end{equation}
\begin{equation}\label{eq:A in t* r}
A=\frac{qM}{r}dt^*,
\end{equation}
where $g_{S^2}$ is the metric on the unit $2$-sphere.

\begin{Remark}[Adding $\mathcal{H}^-$ to the Manifold]
	Normally, the Reissner--Nordstr\"om manifold includes $\mathcal{H}^-$, and in the sub-extremal case the bifurcation sphere $\mathcal{B}$. However, as we will be considering the exterior of a collapsing dust cloud, we will not need $\mathcal{H}^-$. Thus, we will not concern ourselves with the intricacies of attaching $\mathcal{H}^-$ and $\mathcal{B}$ to $\mathcal{M}_{RN}$.
\end{Remark}

We now proceed to calculate the path moved by a radially moving charged test particle, with charge density $\tilde{q}$. The motion of this particle extremises the following action:

\begin{align}
S&=\int_{r=r_b}Ld\tau=m\int_{r=r_b}\frac{1}{2}g_{ab}v^av^b-\tilde{q}v^aA_ad\tau\\\nonumber
&=m\int_{r=r_b}\left(\frac{1}{2}\left(-\left(1-\frac{2M}{r}+\frac{q^2 M^2}{r^2}\right)\left(\frac{dt^*}{d\tau}\right)^2+2\left(\frac{2M}{r}-\frac{q^2 M^2}{r^2}\right)\frac{dt^*}{d\tau}\frac{dr}{d\tau}+\left(1+\frac{2M}{r}-\frac{q^2 M^2}{r^2}\right)\left(\frac{dr}{d\tau}\right)^2\right)-\frac{q\tilde{q}M}{r}\frac{dt^*}{d\tau}\right)d\tau
\end{align}
for $v^a$ the velocity of the particle with respect to $\tau$, $A$ as defined in \eqref{eq:A Definition}, and $\tau$ the proper time for the particle, i.e.~normalised such that $g_{ab}v^av^b=-1$.

We can then use first integrals of the Euler--Lagrange equations to find constants of the motion.
Firstly, $L$ is independent of explicit $\tau$ dependence, and so $g_{ab}v^av^b$ is constant. By rescaling $\tau$, we choose $g_{ab}v^av^b$ to be $-1$. The second constant we obtain is from $L$ being independent of $t^*$. Thus
\begin{equation}\label{eq:T^*}
T^*=\left(1-\frac{2M}{r}+\frac{q^2 M^2}{r^2}\right)\frac{dt^*}{d\tau}-\left(\frac{2M}{r}-\frac{q^2 M^2}{r^2}\right)\frac{dr}{d\tau}+\frac{q\tilde{q}M}{r}
\end{equation}
is constant.

Using \eqref{eq:T^*} to remove dependence of $g_{ab}v^av^b$ on $\frac{dt^*}{d\tau}$, we obtain
\begin{align}\nonumber
\left(1-\frac{2M}{r}+\frac{q^2 M^2}{r^2}\right)^{-1}\left(T^*-\frac{q\tilde{q}M}{r}\right)^2&=\left(1-\frac{2M}{r}+\frac{q^2M^2}{r^2}\right)\left(\frac{dt^*}{d\tau}\right)^2-2\left(\frac{2M}{r}-\frac{q^2 M^2}{r^2}\right)\frac{dt^*}{d\tau}\frac{dr}{d\tau}+\frac{\left(\frac{2M}{r}-\frac{q^2 M^2}{r^2}\right)^2}{1-\frac{2M}{r}+\frac{q^2 M^2}{r^2}}\left(\frac{dr}{d\tau}\right)^2\\\nonumber
&=-g_{ab}v^av^b+\left(1+\frac{2M}{r}-\frac{q^2 M^2}{r^2}+\frac{\left(\frac{2M}{r}-\frac{q^2 M^2}{r^2}\right)^2}{1-\frac{2M}{r}+\frac{q^2 M^2}{r^2}}\right)\frac{dr}{d\tau}^2\\
&=1+\left(1-\frac{2M}{r}+\frac{q^2 M^2}{r^2}\right)^{-1}\left(\frac{dr}{d\tau}\right)^2,
\end{align}
which rearranges to
\begin{align}\label{eq:dr/dtau}
\left(\frac{dr}{d\tau}\right)^2&=\left(T^*-\frac{q\tilde{q}M}{r}\right)^2-\left(1-\frac{2M}{r}+\frac{q^2 M^2}{r^2}\right)\\\nonumber
&=\left(T^*-\frac{q\tilde{q}M}{r}\right)^2-\left(1-\frac{M}{r}\right)^2+\frac{(1-q^2)M^2}{r^2}\\\nonumber
&=\frac{Mr_+}{r^2}\left(T^*-\frac{Mq\tilde{q}}{r_+}+\frac{(T^*-1)(r-r_+)}{M}\right)\left(T^*-\frac{Mq\tilde{q}}{r_+}+\frac{(T^*+1)(r-r_+)}{M}\right).
\end{align}

From \eqref{eq:T^*} we can see that if this particle's velocity is to be future directed, we require $T^*>0$.

In order for the surface of this dust cloud to cross the event horizon, we require that $\tilde{q}<\frac{T^*r_+}{Mq}$. We can then see that $\frac{dr}{d\tau}\neq 0$ for any $r$, provided $T^*>1$. If $T^*<1$, then $\frac{dr}{d\tau}=0$ for some $r$ (we will discuss if this $r$ is obtained in finite time below).

We can also see, from writing out the statement $g_{ab}v^av^b=-1$, that
\begin{equation}\label{eq:timelike}
\left(-\left(1-\frac{2M}{r}+\frac{q^2 M^2}{r^2}\right)+2\left(\frac{2M}{r}-\frac{q^2 M^2}{r^2}\right)\frac{dr}{dt^*}+\left(1+\frac{2M}{r}-\frac{q^2 M^2}{r^2}\right)\left(\frac{dr}{dt^*}\right)^2\right)\left(\frac{dt^*}{d\tau}\right)^2=-1
\end{equation}
which tells us that $\frac{dr}{dt^*}>-1$.
\subsubsection{$T^*<1$}
We now look at the behaviour of $r_b$ in the case where $T^*<1$. 

If $T^*<1$, then looking at $r\to\infty$ we can see $\frac{dr}{d\tau}$ vanishes at a finite radius, so the matter cloud will tend to that radius, either reaching it at a finite time, or as $t\to-\infty$.

We therefore look at integrating equation \eqref{eq:dr/dtau} to obtain $\tau(r)$, which gives us
\begin{equation}\label{eq:tau(r)}
\tau=\frac{M}{(1-T^{*2})^{3/2}}\left((1-q\tilde{q}T^*)\sin^{-1}\left(\frac{1-{T^*}^2}{A}\frac{r}{M}-\frac{1-q\tilde{q}T^*}{A}\right)-\left(A-\left((1-T^{*2})\frac{r}{M}-(1-q\tilde{q}T^*)\right)^2\right)^\frac{1}{2}\right)
\end{equation}
where $A=\sqrt{(1-q\tilde{q}T^*)^2-q^2(1-\tilde{q}^2)(1-{T^*}^2)}$ is a constant.

Equation \eqref{eq:tau(r)} tells us that in the case $T^*<1$, the matter cloud's radius obtains its limit within a finite (and therefore compact) proper time interval. As $t^*$ is a continuous increasing function of $\tau$, $r_b$ obtains its limit in finite $t^*$ time. We will call this time $t^*_-$. At this point, the curve would collapse back into the black hole, hitting the past event horizon. Therefore, in order to have a collapsing model in the $T^*<1$ case, we will assume that the radius of the matter cloud, $r_b(t^*)$, remains at $r_b(t^*_-)$ for all $t^*\leq t^*_-$.

\subsubsection{$T^*\geq 1$}
Here we have that our matter cloud radius tends to $\infty$, as $\tau\to-\infty$. Thus the main part we will need to concern ourselves with is what happens to the surface of the matter cloud as $\tau\to-\infty$, $r_b\to\infty$. Equations \eqref{eq:dr/dtau} and \eqref{eq:T^*} give us that
\begin{equation}\label{eq:dr/dtau limit}
\dot{r_b}:=\frac{dr}{dt^*}=\frac{\frac{dr}{d\tau}}{\frac{dt^*}{d\tau}}\to-\frac{\sqrt{{T^*}^2-1}}{T^*}=:-a \qquad \text{as}\quad t^*\to-\infty
\end{equation}
where we will refer to $a\in[0,1)$ as the asymptotic speed of the surface of the matter cloud. Note that in the case $T^*=1$, we obtain that $\dot{r_b}\to 0$ as $t^*\to-\infty$. In this case, 

\subsection{Definition of RNOS Manifold and Global Coordinates}\label{Manifold Definition}

The RNOS models are defined as a class of collapsing spacetimes, with parameters $M\geq 0$, $q\in[-1,1]$, and a $H^2_{loc}$ function $r_b:(-\infty,t^*_c]\to[0,\infty)$ with constraints \eqref{eq:rbnonincreasing}-\eqref{eq:rbtimelike}. The topologies of the underlying manifolds are all given by global coordinates:
\begin{equation}\label{eq:Manifold}
\mathcal{M}=\bigcup_{t^*\in\R}\{t^*\}\times[\tilde{r}_b(t^*),\infty)\times S^2\subset \mathcal{M}_{RN}.
\end{equation}
where the range of the second coordinate, $r$ depends on the first coordinate $t^*$. Then we have metric
\begin{align}\label{eq:metric}
g=-\left(1-\frac{2M}{r}+\frac{q^2 M^2}{r^2}\right){dt^*}^2+2\left(\frac{2M}{r}-\frac{q^2 M^2}{r^2}\right)&dt^*dr+\left(1+\frac{2M}{r}-\frac{q^2 M^2}{r^2}\right)dr^2+r^2g_{S^2}\\\nonumber
t^*\in\R\qquad& r\in[\tilde{r}_b(t^*),\infty)
\end{align}
where $g_{S^2}$ is the Euclidean metric on the unit sphere.

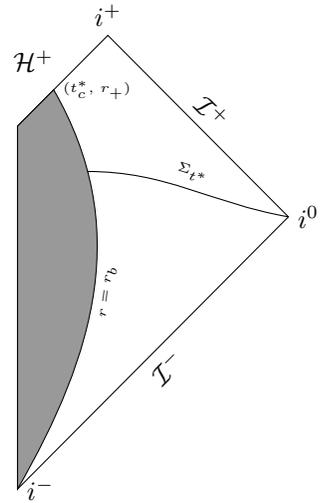
\begin{wrapfigure}{r}{5cm}\label{Fig:Penrose3}
	\vspace{-1cm}
	\begin{tikzpicture}[scale =1.2]
		\node (I)    at ( 0,0) {};
		
		\path 
		(I) +(90:2)  coordinate[label=90:$i^+$]  (Itop)
		+(-90:2) coordinate (Imid)
		+(0:2)   coordinate[label=0:$i^0$] (Iright)
		+(-1,1) coordinate (Ileft)
		+(-0.6,1.4) coordinate[label=0:\tiny ($t^*_c$\text{, }$r_+$)] (BHH)
		+(-1,-3) coordinate[label=0:$i^-$] (Ibot)
		;
		\draw (Ileft) -- 
		node[midway, above left]    {$\mathcal{H}^+$}
		(Itop) --
		node[midway, above, sloped] {$\mathcal{I}^+$}
		(Iright) -- 
		node[midway, below, sloped] {$\mathcal{I}^-$}
		(Ibot) --
		node[midway, above, sloped]    {\small }    
		(Ileft) -- cycle;
		\draw[fill=gray!80] (Ibot) to[out=60, in=-60]
		node[midway, below, sloped] {\tiny $r=r_b$} (BHH)--(Ileft)--cycle;
		\draw (Iright) to[out=170, in=0] node[midway, above, sloped] {\tiny $\Sigma_{t^*}$} (-0.23,0.5);
	\end{tikzpicture}
	\caption{Penrose Diagram of RNOS Model, with spacelike hyper surface $\Sigma_{t^*}$.}\label{fig:PenRef3}
\end{wrapfigure}
We impose the following conditions on $r_b$:
\begin{align}\label{eq:rbnonincreasing}
	\dot{r_b}(t^*):=\frac{dr}{dt^*}&\in(-1,0]\\\label{eq:t^*_cdefinition}
	\exists t^*_c \text{ s.t. }r_b(t^*_c&)=r_+, r_b(t^*)>r_+\quad\forall t^*<t^*_c\\\label{eq:rbtimelike}
	(1,\dot{r}_b(t^*),0,0)&\in T(\mathcal{M})\text{ is timelike for all } t^*\in (-\infty,t^*_c],
\end{align}
where $r_+$ is the black hole horizon for the Reissner--Nordstr\"om spacetime given by \eqref{eq:r+}. We then define $\tilde{r}_b$ by
\begin{equation}
	\tilde{r}_b:=\begin{cases}
		r_b(t^*)&t^*\leq t^*_c\\
		r_+ & t^*>t^*_c
	\end{cases},
\end{equation}

Note that $\tilde{r}_b(t^*)$ is not a differentiable function of $t^*$, but $r_b$ is.

We allow 2 possible past asymptotic behaviours for $r_b$. Firstly,
\begin{equation}
\int_{-\infty}^{t^*_c} \vert\dot{r}_b(t^*)\vert+\vert\ddot{r}_b(t^*)\vert dt^*<\infty.
\end{equation}
This is known as the `fixed boundary' case, as it required $r_b(t^*)\to r_-$ as $t^*\to -\infty$ for some $r_-$.

The second allowed past asymptotic behaviour will be referred to as the `expanding boundary' case, and requires:
\begin{align}
\int_{-\infty}^{t^*_c}\frac{1}{r_b(t^*)^2}dt^*&<\infty\\\nonumber
\dot{r}_b(t^*)\in[-1+\epsilon,0]& \text{ for some }\epsilon>0.
\end{align}
This model includes any past boundary condition for which $\dot{r}_b\to a\in(-1,0)$, and also includes the Oppenheimer--Snyder model, as this has $r_b(t^*)\sim(-t^*)^{2/3}$. Note this case requires $r_b\to\infty$ as $t^*\to-\infty$.

The formation of an extremal black hole in finite time is a much discussed topic, see for example \cite{Israel}, \cite{WALD1974548} and more recently \cite{kehle2022gravitational}, and is heavily related to the third law of black hole dynamics. However, this paper will not discuss the formation of these black holes in more detail, and instead just consider the fairly general RNOS models given above.

The RNOS models have the same exterior Penrose diagram as the original Oppenheimer--Snyder model, see Figure \ref{fig:PenRef}, derived in \cite{Mine}, for example.

We will also be using the double null coordinates given by:
\begin{align}\label{eq:udefinition}
u&=t^*-\int_{s=3M}^r\frac{1+\frac{2M}{s}-\frac{q^2M^2}{s^2}}{1-\frac{2M}{s}+\frac{q^2M^2}{s^2}}ds\\\label{eq:vdefinition}
v&=t^*+r\\
\p_u&=\frac{1}{2}\left(1-\frac{2M}{r}+\frac{q^2M^2}{r^2}\right)\left(\p_{t^*}-\p_r\right)\\
\p_v&=\frac{1}{2}\left(\left(1+\frac{2M}{r}-\frac{q^2M^2}{r^2}\right)\p_{t^*}+\left(1-\frac{2M}{r}+\frac{q^2M^2}{r^2}\right)\p_r\right)\\
g&=-\left(1-\frac{2M}{r}+\frac{q^2M^2}{r^2}\right)dudv+r(u,v)^2g_{S^2}.
\end{align}

Finally, we have four linearly independent Killing vector fields in our space time. The timelike Killing field, $\p_{t^*}$ does not preserve the boundary $\{r=r_b(t^*)\}$. However, we have $3$ angular Killing fields, $\{\Omega_i\}_{i=1}^3$ which span all angular derivatives and are tangent to the boundary of the matter cloud. When given in $\theta$, $\varphi$ coordinates, these take the form:
\begin{align}\nonumber
\Omega_1&=\p_{\varphi}\\\label{eq:AngularKilling}
\Omega_2&=\cos\varphi\p_{\theta}-\sin\varphi\cot\theta\p_{\varphi}\\\nonumber
\Omega_3&=-\sin\varphi\p_{\theta}-\cos\varphi\cot\theta\p_{\varphi}.
\end{align}

\section{Notation}
In this paper, we will be using the same notation as \cite{Mine}. 

We will be considering the following hypersurfaces in our manifold, equipped with the stated normals and volume forms. Note these normals will not neccessarily be unit normals, but have been chosen such that divergence theorem can be applied without involving additional factors.
\begin{align}
&\Sigma_{t^*_0}:=\{(t^*,r,\theta,\varphi):r\geq \tilde{r}_b(t^*), t^*=t^*_0\} &&dV=r^2drd\omega &dn=-dt^*\\
&\Sigma_{u_0}:=\{(t^*,r,\theta,\varphi):r\geq \tilde{r}_b(t^*), u(t^*,r)=u_0\} &&dV=\frac{1}{2}\left(1-\frac{2M}{r}+\frac{q^2M^2}{r^2}\right)r^2dvd\omega &dn=-du\\
&\Sigma_{v_0}:=\{(t^*,r,\theta,\varphi):r\geq \tilde{r}_b(t^*), v(t^*,r)=v_0\} &&dV=\frac{1}{2}\left(1-\frac{2M}{r}+\frac{q^2M^2}{r^2}\right)r^2dud\omega &dn=-dv\\
&S_{[t^*_0,t^*_1]}=\{(t^*,\tilde{r}_b(t^*),\theta,\varphi)\text{ s.t. }t^*\in[t^*_0,t^*_1]\}&& dV=r^2dt^*d\omega& dn=d\rho:=dr-\dot{\tilde{r}}_bdt^*,
\end{align}
where $d\omega$ is the Euclidean volume form on the unit sphere \textit{i.e.}
\begin{equation}
d\omega=\sin\theta d\theta d\varphi.
\end{equation}

We define future/past null infinity by:
\begin{equation}
\mathcal{I}^+:=\R\times S^2\qquad dV=dud\omega\qquad\mathcal{I}^-:=\R\times S^2\qquad dV=dvd\omega.
\end{equation}
Past null infinity is viewed as the limit of $\Sigma_{u_0}$ as $u_0\to\infty$. For an appropriate function $f(u,v,\theta,\varphi)$, we will define the function ``evaluated on $\mathcal{I}^+$" to be
\begin{equation}
f(v,\theta,\varphi)\vert_{\mathcal{I}^-}:=\lim_{u\to-\infty}f(u,v,\theta,\varphi).
\end{equation}

Similarly, $\mathcal{I}^+$ is considered to be the limit of $\Sigma_{v_0}$ as $v_0\to\infty$. For an appropriate function $f(u,v,\theta,\varphi)$, we will define the function ``evaluated on $\mathcal{I}^+$" to be
\begin{equation}
f(u,\theta,\varphi):=\lim_{v\to\infty}f(u,v,\theta,\varphi).
\end{equation}

From here onwards, any surface integral that is left without a volume form will be assumed to have the relevant volume form listed above, and all space-time integrals will be assumed to have the usual volume form $\sqrt{-det(g)}$.

We will be considering solutions of \eqref{eq:wave} which vanish on the surface $r=r_b(t^*)$ (in a trace sense). We will generally be considering these solutions to arise from initial data on a spacelike surface. Initial data will consist of imposing the value of the solution and its normal derivative, with both smooth and compactly supported. 

We will then consider the following seminorms of a spacetime function $f$ on any given submanifold $\Sigma\subset\mathcal{M}$, given by:

\begin{equation}\label{eq:L2}
\Vert f\Vert^2_{L^2(\Sigma)}=\int_{\Sigma}\vert f\vert^2dV.
\end{equation}

We will also define the $\dot{H}^1$ norm as:
\begin{align}
\Vert f\Vert_{\dot{H}^1(\Sigma_{t^*_0})}^2&:=\int_{\Sigma_{t^*_0}}\vert\p_{t^*}f\vert^2+\vert\p_{r}f\vert^2+\frac{1}{r^2}\Vert\mathring{\slashed\nabla{f}}\Vert^2 dV\\
\Vert f\Vert_{\dot{H}^1(\Sigma_{u_0})}^2&:=\int_{\Sigma_{u_0}}\frac{\vert\p_{v}f\vert^2}{\left(1-\frac{2M}{r}+\frac{q^2M^2}{r^2}\right)^2}+\frac{1}{r^2}\Vert\mathring{\slashed\nabla{f}}\Vert^2 dV\\
\Vert f\Vert_{\dot{H}^1(\Sigma_{v_0})}^2&:=\int_{\Sigma_{v_0}}\frac{\vert\p_{u}f\vert^2}{\left(1-\frac{2M}{r}+\frac{q^2M^2}{r^2}\right)^2}+\frac{1}{r^2}\Vert\mathring{\slashed\nabla{f}}\Vert^2 dV,
\end{align}
where $\mathring{\slashed\nabla}$ is the induced gradient on the unit sphere. This is a tensor on the unit sphere, and we define the norm of such a tensor by
\begin{equation}
\Vert T\Vert^2=\sum_{a_1,a_2,...a_m=1}^{n}\vert T_{a_1,a_2,...a_n}\vert^2
\end{equation}
for $T$ an $m$ tensor on $S^n$, in any orthonormal basis tangent to the sphere at that point.

Note that we have not yet defined the spaces for which the $\dot{H}^1$ norms will actually be norms. 

Let $C_0^{\infty}(S)$ be the space of compactly supported functions on surface $S$, which vanish on $\{r=r_b(t^*)\}\cap S$. We will define the $\dot{H}^1(\Sigma_{t^*_0})$ norm on a pair of functions $\phi_0,\phi_1\in C^{\infty}_0(\Sigma_{t^*_0})$ as follows:

\begin{align}
\Vert (\phi_0,\phi_1)\Vert_{\dot{H}^1(\Sigma_{t^*_0})}:=\Vert\phi\Vert_{\dot{H}^1(\Sigma_{t^*_0})}\qquad \text{for any }\phi\text{ s.t. }(\phi\vert_{\Sigma_{t^*_0}},\p_{t^*}\phi\vert_{\Sigma_{t^*_0}})=(\phi_0,\phi_1).
\end{align}

We similarly define the $\dot{H}^1(\Sigma_{u_0})$ and $\dot{H}^1(\Sigma_{v_0})$ on $\phi_0\in C^{\infty}_0(\Sigma_{u_0,v_0})$ as follows:

\begin{align}
\Vert \phi_0\Vert_{\dot{H}^1(\Sigma_{u_0,v_0})}:=\Vert\phi\Vert_{\dot{H}^1(\Sigma_{u_0,v_0})}\qquad \text{for any }\phi\text{ s.t. }\phi\vert_{\Sigma_{u_0,v_0}}=\phi_0.
\end{align}

We will also need to consider what functions we will be working with. For this, we will be using the same notation as \cite{KerrScatter,Mine}. We first need to look at the notions of energy momentum tensors and energy currents (note this energy momentum tensor will be expressed as $T$, and is different from $\boldsymbol{T}$ in \eqref{eq:EE}).
\begin{align}
T_{\mu\nu}(\phi)&=\nabla_\mu\phi\nabla_\nu\phi-\frac{1}{2}g_{\mu\nu}\nabla^\rho\phi\nabla_\rho\phi\\
J^X_{\mu}&=X^\nu T_{\mu\nu}\\
K^X&=\nabla^\mu J^X_\mu\\
\label{eq:modcurrent}
J^{X,w}_\mu&=X^\nu T_{\mu\nu}+w\nabla_\mu(\phi^2)-\phi^2\nabla_\mu w\\
K^{X,w}&=\nabla^\nu J^{X,w}_\nu=K^X+2w\nabla_\mu\phi\nabla^\mu\phi-\phi^2\Box_gw\\
\label{eq:energydef}X\text{-energy}(\phi,S)&=\int_Sdn(J^X).
\end{align}
Here, $dn$ is the normal to $S$. It should be noted that applications of divergence theorem do not introduce any additional factors with our choice of volume form and normal, i.e.
\begin{equation}\label{eq:EnergyIdentity}
	\int_{t^*\in[t^*_0,t^*_1]}K^{X,\omega}=-\int_{\Sigma_{t^*_1}}dn(J^{X,\omega})+\int_{\Sigma_{t^*_0}}dn(J^{X,\omega})-\int_{S_{[t^*_0,t^*_1]}}dn(J^{X,\omega}),
\end{equation}
with similar equations holding for $\Sigma_{u,v}$.

For any $T_{\mu\nu}$ obeying the dominant energy condition, $X$ future pointing and causal, and $S$ spacelike, then the $X$-energy is non-negative. 

For any pair of functions, $\phi_0,\phi_1\in C^\infty_0(\Sigma_{t^*_0})$, and $X$ a causal, future pointing vector, we define the $X$ norm by
\begin{equation}
\Vert (\phi_0,\phi_1)\Vert^2_{X,\Sigma_{t^*_0}}:=X\text{-energy}(\phi,\Sigma_{t^*_0})\qquad \text{for any }\phi\text{ s.t. }(\phi\vert_{\Sigma_{t^*_0}},\p_{t^*}\phi\vert_{\Sigma_{t^*_0}})=(\phi_0,\phi_1).
\end{equation}

We similarly define for $\phi_0\in C^\infty_0(\Sigma_{u_0,v_0})$
\begin{equation}
\Vert\phi_0\Vert_{X,\Sigma_{u_0,v_0}}^2:=X\text{-energy}(\phi,\Sigma_{t^*_0})\qquad \text{for any }\phi\text{ s.t. }\phi\vert_{\Sigma_{t^*_0}}=\phi_0.
\end{equation}

Note that for any causal, future pointing $X$ which coincides with the timelike Killing vector field $\p_{t^*}$ in a neighbourhood of $\mathcal{I}^\pm$, we have that the $X$ norm is Lipschitz equivalent to the $\dot{H}^1$ norm.

For causal timelike vector $X$, we define the following function spaces
\begin{align}
\mathcal{E}_{\Sigma_{t^*_0}}^X&:=Cl_{X,\Sigma_{t^*_0}}(C^\infty_0(\Sigma_{t^*_0})\times C^\infty_0(\Sigma_{t^*_0}))\\
\mathcal{E}_{\Sigma_{u_0,v_0}}^X&:=Cl_{X,\Sigma_{u_0,v_0}}(C^\infty_0(\Sigma_{u_0,v_0}))
\end{align}
where these closures are in $H^1_{loc}$ with respect to the subscripted norms.

For $\psi_0\in C^\infty_0(\mathcal{I}^{\pm})$, we define
\begin{align}
\Vert\psi_0\Vert^2_{\p_{t^*},\mathcal{I}^+}&:=\int_{\mathcal{I}^+}\vert\p_{v}\psi_0\vert^2dvd\omega\\
\Vert\psi_0\Vert^2_{\p_{t^*},\mathcal{I}^-}&:=\int_{\mathcal{I}^-}\vert\p_{u}\psi_0\vert^2dud\omega.
\end{align}

Finally, we define the energy spaces $\mathcal{E}^{\p_{t^*}}_{\mathcal{I}^\pm}$ by
\begin{equation}
\mathcal{E}^{\p_{t^*}}_{\mathcal{I}^\pm}:=Cl_{\p_{t^*},\mathcal{I}^\pm}(C^\infty_0(\mathcal{I}^\pm)).
\end{equation}

\section{Existence and Uniqueness of Solutions}
Given smooth compactly supported initial data $\phi=\phi_0$, $\p_{t^*}
\phi=\phi_1$ on $\Sigma_{t^*_0}$, $t^*_0<t^*_c$, we have that there exists a unique smooth solution compactly supported on every $\Sigma_{t^*}$.

Thus when proving boundedness or decay results, we may assume that our solution has sufficiently many derivatives and that weighted integrals with weights growing in $r$ converge. Then we can generalise results to all $\dot{H}^1$ functions using that compactly supported smooth functions are dense within $\dot{H}^1$ functions.

\begin{Theorem}[Existence of Smooth Solutions]\label{Thm:Existence}
	Let $\phi_0$ and $\phi_1$ smooth, compactly supported functions on $\Sigma_{t^*_0}$, $t^*_0\leq t^*_c$, such that $\phi_0(r_b(t^*_0),\theta,\varphi)=0$ and $\phi_1(r_b(t^*_0),\theta,\varphi)+\dot{r}_b(t^*_0)\p_r\phi_0(r_b(t^*_0),\theta,\varphi)=0$. Then there exists a $\phi\in C^\infty(\mathcal{M})$ with $\phi\vert_{\Sigma{t^*}}\in C^{\infty}_0(\Sigma_{t^*})$ for all $t^*\in\R$, such that
	\begin{align}
	\Box_g\phi&=0\\\label{eq:BoundaryConditions}
	\phi(t^*,r_b(t^*),\theta,\varphi)&=0 \quad\forall t^*\leq t^*_c,\quad (\theta,\varphi)\in S^2\\
	(\phi,\p_{t^*}\phi)\vert_{\Sigma_{t^*_0}}&=(\phi_0,\phi_1).
	\end{align}
\end{Theorem}

\begin{proof}
	For a proof, one can follow the proof of Theorem $5.1$ in \cite{Mine} almost exactly.
\end{proof}

\section{Energy Boundedness}
In this section we work towards proving Theorem \ref{Thm:VagueBound}, as stated in the overview. We will prove this in two sections. We will first prove that going forwards we have a uniform bound on the $\dot{H}^1$ norm, i.e.~there exists a constant $C(\mathcal{M})$ such that
\begin{equation}\label{eq:ForwardBound}
\Vert\phi\Vert_{\dot{H}^1(\Sigma_{t^*_1})}\leq C\Vert\phi\Vert_{\dot{H}^1(\Sigma_{t^*_0})}\quad\forall t^*_1\geq t^*_0.
\end{equation}

In the second section we will prove the analogous statement going backwards in time:
\begin{equation}\label{eq:BackwardBound}
\Vert\phi\Vert_{\dot{H}^1(\Sigma_{t^*_0})}\leq C\Vert\phi\Vert_{\dot{H}^1(\Sigma_{t^*_1})}\quad\forall t^*_0\leq t^*_1\leq t^*_c.
\end{equation}
Note the backwards in time version includes a condition on $t^*_1\leq t^*_c$, as, were $t^*_1>t^*_c$, we can lose arbitrarily large amounts of energy across the event horizon.

From here on in this paper, when we say \emph{solution}, unless stated otherwise, we mean $\phi$ which has finite $\dot{H}^1(\Sigma_{t^*})$ norm for all $t^*$, and is a solution of \eqref{eq:wave} in a distributional sense, i.e.
\begin{equation}
\int_\mathcal{M} g^{ab}\p_bf\p_a\phi=0 \quad \forall f\in C^\infty_0(\mathcal{M}).
\end{equation}
Again, note that smooth compactly supported solutions of \eqref{eq:wave} are dense within these functions with respect to the $\dot{H}^1(\Sigma_{t^*})$ norm. The methods in this section will closely follow \cite{Mine}.
\subsection{Finite in Time Boundedness}
We begin by proving a local in time bound on solutions of \eqref{eq:wave}.

\begin{Theorem}[Finite in Time Energy Bound]\label{Thm:FiniteBound}
	Given an RNOS model $\mathcal{M}$, with associated $M, q, r_b$, $\phi$ a solution of the wave equation \eqref{eq:wave} with boundary conditions \eqref{eq:BoundaryConditions}, and a time interval $t^*_0\leq t^*_1\leq t^*_c$, we have that there exists a constant $C=C(\mathcal{M},t^*_0)>0$ such that
	\begin{equation}
	C^{-1}\Vert\phi\Vert_{\dot{H}^1(\Sigma_{t^*_0})}\leq\Vert\phi\Vert_{\dot{H}^1(\Sigma_{t^*_1})}\leq C\Vert\phi\Vert_{\dot{H}^1(\Sigma_{t^*_0})}
	\end{equation}
\end{Theorem}
\begin{proof}
	We start by proving the result for $\phi$ compactly supported on each $\Sigma_{t^*}$, as then the result can be extended to all $\dot{H}^1$ functions by an easy density argument. We choose a vector field which is everywhere timelike, including on the surface of the matter cloud. We also choose this vector field to be tangent to the surface of the matter cloud. For example
	\begin{equation}
	X=\p_{t^*}+\dot{r_b}(t^*)\p_r.
	\end{equation}
	
	Then we have that
	\begin{align}\nonumber
	-dt^*(J^X)=\frac{1}{2}\Bigg(&\left(1+\frac{2M}{r}-\frac{q^2M^2}{r^2}\right)(\p_{t^*}\phi)^2+2\left(1+\frac{2M}{r}-\frac{q^2M^2}{r^2}\right)\dot{r_b}(t^*)\p_{t^*}\phi\p_r\phi\\
	&+\left(1-\frac{2M}{r}+\frac{q^2M^2}{r^2}+\frac{2M}{r}\left(2-\frac{q^2M}{r}\right)\vert\dot{r_b}(t^*)\vert\right)(\p_r\phi)^2+\frac{1}{r^2}\vert\mathring{\slashed\nabla}\phi\vert^2\Bigg).
	\end{align}
	
	Note, in every RNOS model, when the matter cloud crosses $r=r_+$, $\dot{r_b}(t^*)\neq 0$, and as $\dot{r_b}(t^*)\in(-1,0]$, we have that there exists a time independent constant $A=A(\mathcal{M})$ such that
	\begin{equation}\label{eq:FiniteFluxSim}
	A^{-1}\Vert\phi\Vert_{\dot{H}^1(\Sigma_{t^*})}^2\leq-\int_{\Sigma_{t^*}}dt^*(J^X)\leq A\Vert\phi\Vert_{\dot{H}^1(\Sigma_{t^*})}^2
	\end{equation}
	for all $t^*\leq t^*_c$.
	
	Then we look at the energy current through the surface of the matter cloud
	\begin{equation}
	d\rho(J^X)=0
	\end{equation}
	once we notice that on the surface of the matter cloud $X^\nu\nabla_\nu\phi=0$ and $d\rho(X)=0$ for $d\rho$ the normal to the surface of the matter cloud.
	
	If we then calculate $K^X$, we get:
	\begin{align}\nonumber
	\vert K^X\vert=\Bigg\vert&\left(1+\frac{M}{r}\right)\frac{\dot{r_b}(t^*)}{r}(\p_{t^*}\phi)^2-\left(\frac{2M\dot{r_b}(t^*)}{r^2}+\left(1+\frac{2M}{r}-\frac{q^2M^2}{r^2}\right)\ddot{r_b}(t^*)\right)\p_{t^*}\phi\p_r\phi\\
	&-\left(\left(1-\frac{M}{r}\right)\frac{\dot{r_b}(t^*)}{r}-\left(\frac{2M}{r}-\frac{q^2M^2}{r^2}\right)\ddot{r_b}(t^*)\right)(\p_r\phi)^2\Bigg\vert\leq B\left(\vert\dot{r}_b\vert+\vert\ddot{r}_b\vert\right)\dot\Vert\phi\Vert_{\dot{H}^1(\Sigma_{t^*})}^2.
	\end{align}
	
	Define
	\begin{equation}
	f(t^*):=-\int_{\Sigma_{t^*}}dt^*(J^X)=\Vert\phi\Vert_{X,\Sigma_{t^*}}^2.
	\end{equation}
	
	Now, if we integrate $K^X$ in the region $t^*\in[t^*_0,t^*_1]$ and apply \eqref{eq:EnergyIdentity}, we obtain
	\begin{equation}\label{eq:FiniteGronwall}
		f(t^*_1)-B\int_{t^*_0}^{t^*_1}\left(\vert\dot{r}_b\vert+\vert\ddot{r}_b\vert\right)f(t^*)dt^*\leq f(t^*_0)\leq f(t^*_1)+B\int_{t^*_0}^{t^*_1}\left(\vert\dot{r}_b\vert+\vert\ddot{r}_b\vert\right)f(t^*)dt^*.
	\end{equation}
	Then an application of Gronwall's Inequality to each of the inequalities in \eqref{eq:FiniteGronwall}, and applying \eqref{eq:FiniteFluxSim} to the result gives us that
	\begin{equation}
		\left(Ae^{B\int_{t^*_0}^{t^*_1}\left(\vert\dot{r}_b\vert+\vert\ddot{r}_b\vert\right)dt^*}\right)^{-1}\Vert\phi\Vert_{\dot{H}^1(\Sigma_{t^*_0})}^2\leq\Vert\phi\Vert_{\dot{H}^1(\Sigma_{t^*_1})}^2\leq \left(Ae^{B\int_{t^*_0}^{t^*_1}\left(\vert\dot{r}_b\vert+\vert\ddot{r}_b\vert\right)dt^*}\right)\Vert\phi\Vert_{\dot{H}^1(\Sigma_{t^*_0})}^2.
	\end{equation}
	Letting $C^2=Ae^{B\int_{t^*_0}^{t^*_c}\left(\vert\dot{r}_b\vert+\vert\ddot{r}_b\vert\right)dt^*}$ gives us the required result.
\end{proof}
\subsection{Fixed Boundary Case}
We now prove a uniform boundedness result for the case $r_b$ is constant for $t^*\leq t^*_-$.:
\begin{Theorem}[Uniform in Time Energy Bound for the Fixed Boundary Case]\label{Thm:T^*<1}
Given an RNOS model $\mathcal{M}$ and associated $M$, $q$ and $r_b$, with $r_b(t^*)$ constant for $t^*\leq t^*_-$, and $\phi$ a solution of the wave equation \eqref{eq:wave} with boundary conditions \eqref{eq:BoundaryConditions}, we have that there exists a constant $C=C(\mathcal{M})>0$ such that
\begin{equation}
C^{-1}\Vert\phi\Vert_{\dot{H}^1(\Sigma_{t^*_0})}\leq\Vert\phi\Vert_{\dot{H}^1(\Sigma_{t^*_1})}\leq C\Vert\phi\Vert_{\dot{H}^1(\Sigma_{t^*_0})} \quad \forall t^*_0\leq t^*_1\leq t^*_c.
\end{equation}
\end{Theorem}
\begin{proof}
	This proof is identical to the proof for Theorem \ref{Thm:FiniteBound}. By the definition of the fixed boundary case, the integral $\int_{-\infty}^{t^*_c}\left(\vert\dot{r}_b\vert+\vert\ddot{r}_b\vert\right)$ converges. Thus, the constant given by Theorem \ref{Thm:FiniteBound} is actually a uniform in time bound for all $t^*_0<t^*_-$, i.e.~take 
	\begin{equation}
	C^2=Ae^{B\int_{-\infty}^{t^*_c}\left(\vert\dot{r}_b\vert+\vert\ddot{r}_b\vert\right)dt^*}.
	\end{equation}
\end{proof}
\subsection{Expanding Boundary Case}
The expanding boundary case is much more difficult than the fixed boundary case, so we will break this boundedness result into two different Theorems. We will start with the forward bound:
\begin{Theorem}[Uniform Forward in Time Energy Bound for the Expanding Boundary Case]\label{Thm:T^*>1}
	Given an RNOS model $\mathcal{M}$ with associated $M,q,r_b$ with $r_b(t^*)\to\infty$ as $t^*\to-\infty$, and a solution $\phi$ of the wave equation \eqref{eq:wave} with boundary conditions \eqref{eq:BoundaryConditions}, we have that there exists a constant $C=C(\mathcal{M})>0$ such that
	\begin{equation}
	\Vert\phi\Vert_{\dot{H}^1(\Sigma_{t^*_1})}\leq C\Vert\phi\Vert_{\dot{H}^1(\Sigma_{t^*_0})} \quad \forall t^*_0\leq t^*_1\leq t^*_c.
	\end{equation}
\end{Theorem}
\begin{proof}
	We will proceed similarly to Theorem \ref{Thm:FiniteBound}, but we will take the vector field
	\begin{equation}
	X=\p_{t^*}.
	\end{equation}
	Then we obtain the following results:
	\begin{align}\label{eq:KT}
	K^X&=0\\\label{eq:dt(JT)}
	-dt^*(J^X)&=\frac{1}{2}\Bigg(\left(1+\frac{2M}{r}-\frac{q^2M^2}{r^2}\right)(\p_{t^*}\phi)^2+\left(1-\frac{2M}{r}+\frac{q^2M^2}{r^2}\right)(\p_r\phi)^2+\frac{1}{r^2}\vert\mathring{\slashed\nabla}\phi\vert^2\Bigg)\\\label{eq:drho(JT)}
	d\rho(J^X)&=-\frac{\dot{r_b}(t^*)}{2}(1+\dot{r_b}(t^*))\left(\left(1-\frac{2M}{r}+\frac{q^2M^2}{r^2}\right)-\left(1+\frac{2M}{r}-\frac{q^2M^2}{r^2}\right)\dot{r_b}(t^*)\right)(\p_r\phi)^2\geq 0,
	\end{align}
	recalling that $\dot{r}_b\in (-1,0]$.
	
	Now if we take an arbitrary $t^*_0<t^*_c$, then in the region $t^*\leq t^*_0$, $r\geq r_b(t^*)\geq r_b(t^*_0)>r_+$. Thus there exists an $\epsilon>0$ such that 
	\begin{equation}
	1-\frac{2M}{r}+\frac{q^2M^2}{r^2}\geq\epsilon.
	\end{equation}
	
	Therefore in the region $t^*\leq t^*_0$
	\begin{equation}
	\epsilon\Vert\phi\Vert_{\dot{H}^1(\Sigma_{t^*})}^2\leq -\int_{\Sigma_{t^*}}dt^*(J^X)\leq\Vert\phi\Vert_{\dot{H}^1(\Sigma_{t^*})}^2.
	\end{equation}
	
	Then, as before, we integrate $K^X$ in a region $t^*\in[t^*_1,t^*_2]$ for $t^*_1\leq t^*_2\leq t^*_0$, and apply \eqref{eq:EnergyIdentity}. Once we note that the boundary term has the correct sign, we have 
	\begin{equation}
	\Vert\phi\Vert_{\dot{H}^1(\Sigma_{t^*_2})}^2\leq \epsilon^{-1}\Vert\phi\Vert_{\dot{H}^1(\Sigma_{t^*_1})}^2.
	\end{equation}
	An application of Theorem \ref{Thm:FiniteBound} will then allow us to extend our bound over the remaining finite interval $[t^*_0,t^*_c]$ to obtain the required result.
\end{proof}

Now we look at obtaining the backward in time bound:
\begin{Theorem}[Uniform Backward in Time Energy Bound for Expanding Boundary Case]\label{Thm:BackwardsBound}
	Given an RNOS model $\mathcal{M}$ with associated $M$, $q$ and $r_b$ with $r_b\to\infty$ as $t^*\to-\infty$, and a solution $\phi$ of the wave equation \eqref{eq:wave} with boundary conditions \eqref{eq:BoundaryConditions}, we have that there exists a constant $C=C(\mathcal{M})>0$ such that
	\begin{equation}
	\Vert\phi\Vert_{\dot{H}^1(\Sigma_{t^*_0})}\leq C\Vert\phi\Vert_{\dot{H}^1(\Sigma_{t^*_1})} \quad \forall t^*_0\leq t^*_1\leq t^*_c.
	\end{equation}
\end{Theorem}
\begin{proof}
	For this proof, we will need to use the modified currents, as defined in \eqref{eq:modcurrent}. Given $\dot{r}_b\geq-1+\epsilon$, let $b\in[1-\epsilon,1)$. Looking in the region $t^*<0$, we will use the vector field and modifier
	\begin{align}
	X&=\p_{t^*}-b\p_r\\
	w&=-\frac{b}{2r}.
	\end{align}
	We then calculate
	\begin{align}\nonumber
	\left\vert \int_{\Sigma_{t^*}}K^{X,w}\right\vert \leq&\int_{\Sigma_{t^*}}\Bigg\vert\frac{bM}{r^2}\left(1-\frac{q^2M}{r}\right)(\p_{t^*}\phi)^2+\frac{bM}{r^2}\left(1-\frac{q^2M}{r}\right)(\p_r\phi)^2-\frac{2bM}{r^2}\left(1-\frac{q^2M}{r}\right)\p_{t^*}\phi\p_r\phi\\\nonumber
	&\qquad-\frac{b}{r^3}\vert\mathring{\slashed\nabla{\phi}}\vert^2-\frac{b}{r^4}\left(1-\frac{q^2M}{r}\right)\phi^2\Bigg\vert\leq \frac{1}{r_b(t^*)^2}\left(\Vert \phi\Vert_{\dot{H}^1(\Sigma_{t^*})}^2+\int_{\Sigma_{t^*}}\frac{\phi^2}{r^2}\right)\\
	d\rho(J^{X,w})=&-\frac{1}{2}\left(b+\dot{r_b}(t^*)\right)(1+\dot{r_b}(t^*))\left(\left(1-\frac{2M}{r}+\frac{q^2M^2}{r^2}\right)-\left(1+\frac{2M}{r}-\frac{q^2M^2}{r^2}\right)\dot{r_b}(t^*)\right)(\p_r\phi)^2\\
	-dt^*(J^{X,w})=&\frac{1}{2}\Bigg(\left(1+\frac{2M}{r}-\frac{q^2M^2}{r^2}\right)(\p_{t^*}\phi)^2-2b\left(1+\frac{2M}{r}-\frac{q^2M^2}{r^2}\right)\p_{t^*}\phi\p_r\phi\\
	&+\left(\left(1-\frac{2M}{r}+\frac{q^2M^2}{r^2}\right)+\frac{2bM}{r}\left(2-\frac{q^2M}{r}\right)\right)(\p_r\phi)^2
	+\frac{1}{r^2}\vert\mathring{\slashed\nabla}\phi\vert^2\\\nonumber
	&+\frac{2bM}{r}\left(2-\frac{q^2M}{r}\right)\frac{\phi}{r}\p_r\phi-2b\left(1+\frac{2M}{r}-\frac{q^2M^2}{r^2}\right)\frac{\phi}{r}\p_{t^*}\phi+\frac{bM}{r}\left(2-\frac{q^2M}{r}\right)\frac{\phi^2}{r^2}\Bigg).
	\end{align}
	
	We can see that $d\rho(J^X{,w})\geq 0$, for sufficiently negative $t^*$, since $b+\dot{r}_b(t^*)\geq 0$. Next, let us consider $-dt^*(J^{X,w})$. Integrating over $\Sigma_{t^*}$, we obtain:
	\begin{align}
	\nonumber
	-\int_{\Sigma_{t^*}}dt^*(J^{X,w})=\frac{1}{2}\int_{\Sigma_{t^*}}&\Bigg(\left(1+\frac{2M}{r}-\frac{q^2M^2}{r^2}\right)(\p_{t^*}\phi)^2-2b\left(1+\frac{2M}{r}-\frac{q^2M^2}{r^2}\right)\p_{t^*}\phi\p_r\phi\\\nonumber
	&+\left(\left(1-\frac{2M}{r}+\frac{q^2M^2}{r^2}\right)+\frac{2bM}{r}\left(2-\frac{q^2M}{r}\right)\right)(\p_r\phi)^2
	+\frac{1}{r^2}\vert\mathring{\slashed\nabla}\phi\vert^2\\\nonumber
	&+\frac{2bM}{r}\left(2-\frac{q^2M}{r}\right)\frac{\phi}{r}\p_r\phi-2b\left(1+\frac{2M}{r}-\frac{q^2M^2}{r^2}\right)\frac{\phi}{r}\p_{t^*}\phi+\frac{bM}{r}\left(2-\frac{q^2M}{r}\right)\frac{\phi^2}{r^2}\Bigg)\\\nonumber
	=\frac{1}{2}\int_{\Sigma_{t^*}}&\Bigg(\left(1+\frac{2M}{r}-\frac{q^2M^2}{r^2}\right)(1-b)(\p_{t^*}\phi)^2+b\left(1+\frac{2M}{r}-\frac{q^2M^2}{r^2}\right)\left(\left(\p_{t^*}\phi-\p_r\phi-\frac{\phi}{r}\right)^2-2\frac{\phi}{r}\p_r\phi\right)\\
	&+\left(1-\frac{2M}{r}+\frac{q^2M^2}{r^2}\right)(1-b)(\p_r\phi)^2
	+\frac{1}{r^2}\vert\mathring{\slashed\nabla}\phi\vert^2+\frac{bM}{r}\left(2-\frac{q^2M}{r}\right)\left(\frac{\phi}{r}+\p_r\phi\right)^2\Bigg)\\\nonumber
	=\frac{1}{2}\int_{\Sigma_{t^*}}&\Bigg(\left(1+\frac{2M}{r}-\frac{q^2M^2}{r^2}\right)(1-b)(\p_{t^*}\phi)^2+b\left(1+\frac{2M}{r}-\frac{q^2M^2}{r^2}\right)\left(\p_{t^*}\phi-\p_r\phi-\frac{\phi}{r}\right)^2\\\nonumber
	&+b\left(1+\frac{q^2M^2}{r^2}\right)\frac{\phi^2}{r^2}+\left(1-\frac{2M}{r}+\frac{q^2M^2}{r^2}\right)(1-b)(\p_r\phi)^2
	+\frac{1}{r^2}\vert\mathring{\slashed\nabla}\phi\vert^2\\\nonumber
	&+\frac{bM}{r}\left(2-\frac{q^2M}{r}\right)\left(\frac{\phi}{r}+\p_r\phi\right)^2\Bigg).
	\end{align}
	
	We then note a version of Hardy's inequality. If $h$ is a differentiable function of one variable, with $h(0)=0$, then
	\begin{equation}\label{eq:Hardy's}
	\exists C>0\ s.t.\ \int_{\Sigma_{t^*}}\left(\frac{h(r)}{r}\right)^2\leq C\int_{\Sigma_{t^*}} \left(\p_rh(r)\right)^2.
	\end{equation}
	
	Using \eqref{eq:Hardy's}, we have that there exists a $t^*$ independent constant $A$ such that
	\begin{align}
	0\leq \int_{\Sigma_{t^*}}\Bigg(b\left(1+\frac{2M}{r}-\frac{q^2M^2}{r^2}\right)\left(\p_{t^*}\phi-\p_r\phi-\frac{\phi}{r}\right)^2+b\left(1+\frac{q^2M^2}{r^2}\right)\frac{\phi^2}{r^2}+	\frac{1}{r^2}\vert\mathring{\slashed\nabla}\phi\vert^2&\\\nonumber
	+\frac{bM}{r}\left(2-\frac{q^2M}{r}\right)\left(\frac{\phi}{r}+\p_r\phi\right)^2&\Bigg)\leq A\Vert\phi\Vert_{\dot{H}^1(\Sigma_{t^*})}^2
	\end{align}
	
	Note $b\leq 1$. Thus, provided we are away from $t^*_c$, there exists a $t^*$ independent $\epsilon$ such that
	\begin{equation}\label{eq:TfluxisH1}
	\epsilon\Vert\phi\Vert_{\dot{H}^1(\Sigma_{t^*})}^2\leq -\int_{\Sigma_{t^*}}dt^*(J^{X,w})\leq \epsilon^{-1}\Vert\phi\Vert_{\dot{H}^1(\Sigma_{t^*})}^2.
	\end{equation}
	
	If we again let $f(t^*):=-\int_{\Sigma_{t^*}}dt^*(J^{X,w})$, we can apply \eqref{eq:EnergyIdentity} to see that
	\begin{equation}
		f(t^*_0)\leq f(t^*_1)+A\int_{t^*_0}^{t^*_1}\frac{f(t^*)}{r_b(t^*)^2}dt^*.
	\end{equation}

	Thus we can apply Gronwall's Inequality to $f$, along with the condition on the expanding boundary case that $\int r_b^{-2}(t^*)dt^*$ converges to obtain
	\begin{equation}
		f(t^*)\leq f(t^*_0)e^{A\int^{t^*_0}_{-\infty}\frac{1}{r_b(t^*)^2}dt^*},
	\end{equation}
	for all $t^*<t^*_0$. Applying \eqref{eq:TfluxisH1}, we obtain
	\begin{equation}\label{eq:AlmostUniform}
	\Vert\phi\Vert_{\dot{H}^1(\Sigma_{t^*_1})}^2\leq A\Vert\phi\Vert_{\dot{H}^1(\Sigma_{t^*_0})}^2\quad \forall t^*_1\leq t^*_0 \leq t^*_2.
	\end{equation}
	Combining \eqref{eq:AlmostUniform} with Theorem \ref{Thm:FiniteBound} for the interval $[t^*_0,t^*_c]$ gives us the final result.
\end{proof}

\section{The Scattering Map}
We now consider bounds on the radiation fields. We will be considering the maps $\mathcal{G}^+$ and $\mathcal{F}^-$, which take data from $\Sigma_{t^*_c}$ to data on $\mathcal{I}^+$ and $\mathcal{I}^-$ respectively. We will also consider their inverses (where defined), $\mathcal{G}^-$ and $\mathcal{F}^+$, which take data from $\mathcal{I}^+$ and $\mathcal{I}^-$ respectively to $\Sigma_{t^*_c}$. We will look at obtaining boundedness or non-boundedness for these.

Finally, we will define the scattering map, $\mathcal{S}^+:=\mathcal{G}^+\circ\mathcal{F}^+$, and consider boundedness results for this.

\subsection{Existence of Radiation Fields}
To look at these maps, we will first need a definition of radiation field. We will then need to show it exists for all finite energy solutions of the wave equation.
\begin{Proposition}[Existence of the Backwards Radiation Field]\label{Prop:RadFieldExist}
	Given $\phi$ a solution to the wave equation \eqref{eq:wave} with boundary conditions \eqref{eq:BoundaryConditions}, there exist $\psi_{+,-}$ such that
	\begin{equation}
	r(u,v)\phi(u,v,\theta,\varphi)\xrightarrow[u\to-\infty]{H^1_{loc}} \psi_-(v,\theta,\varphi)
	\end{equation}
	\begin{equation}
	r(u,v)\phi(u,v,\theta,\varphi)\xrightarrow[v\to\infty]{H^1_{loc}} \psi_+(u,\theta,\varphi).
	\end{equation}
\end{Proposition}
\begin{proof}
	This existence has been done many times before, see for example \cite{Moschidis,Mine}.
\end{proof}

\subsection{Backwards Scattering from $\Sigma_{t^*_c}$}\label{Sec:Backwards Scattering}
Now we have existence of the radiation field, we define the following map:
\begin{align}
\mathcal{F}^-:\ &\ \ \ \ \ \  \mathcal{E}^X_{\Sigma_{t^*_c}}\ \ \ \ \ \ \ \  \longrightarrow \ \ \ \mathcal{F}^-\left(\mathcal{E}^X_{\Sigma_{t^*_c}}\right)\subset H^1_{loc}(\mathcal{I}^-)\\\nonumber
&(\phi|_{\Sigma_{t^*_c}},\p_{t^*}\phi|_{\Sigma_{t^*_c}})\mapsto \ \ \ \psi_-
\end{align}
\begin{wrapfigure}{r}{5cm}
	\begin{tikzpicture}[scale =1.2]
	\node (I)    at ( 0,0) {};
	\node (D) 	at (0.5,-0.5) {$D_{v_1}$};
	\path 
	(I) +(90:2)  coordinate[label=90:$i^+$]  (Itop)
	+(-90:2) coordinate (Imid)
	+(0:2)   coordinate[label=0:$i^0$] (Iright)
	+(-1,1) coordinate (Ileft)
	+(-0.6,1.4) coordinate (BHH)
	+(-1,-3) coordinate[label=0:$i^-$] (Ibot)
	;
	\draw (Ileft) -- 
	node[midway, above left]    {$\mathcal{H}^+$}
	(Itop) --
	node[midway, above, sloped] {$\mathcal{I}^+$}
	(Iright) -- 
	node[midway, below, sloped] {$\mathcal{I}^-$}
	(Ibot) --
	node[midway, above, sloped]    {\small }    
	(Ileft) -- cycle;
	\draw[fill=gray!80] (Ibot) to[out=60, in=-60]
	node[midway, below, sloped] {\tiny $r=r_b$} (BHH)--(Ileft)--cycle;
	\draw (Iright) to[out=170, in=0] node[midway, above, sloped] {\tiny $\Sigma_{t^*_2}$} (-0.23,0.5);
	\draw (1.48,0.12) to node[midway, below, sloped] {\tiny $\Sigma_{v_0}$} (1.8,-0.2);
	\draw (0,-2) to node[midway, above, sloped] {\tiny $\Sigma_{v_1}$} (-0.35,-1.65);
	\end{tikzpicture}
\end{wrapfigure}
where the $\psi_-$ is as defined in Proposition \ref{Prop:RadFieldExist}, and the $X$ is any everywhere timelike vector field (including on the event horizon) which coincides with the timelike Killing vector field $\p_{t^*}$ for sufficiently large $r$. An $X$ with these properties is chosen, so that the $X$ norm is equivalent to the $\dot{H}^1$ norm.

We define the inverse of $\mathcal{F}^-$ (once injectivity is established on the image of $\mathcal{F}^-$) as $\mathcal{F}^+$.
\begin{itemize}
	\item Firstly, we will show $\mathcal{F}^-$ is bounded. (Proposition \ref{Prop:BackScatBound})
	\item Then we will show that $\mathcal{F}^+$, if it can be defined, would be bounded, which gives us that $\mathcal{F}^-$ is injective. (Proposition \ref{Prop:ForwardScatBound})
	\item Finally, we show that $Im(\mathcal{F}^-)$ is dense in $\mathcal{E}^T_{\mathcal{I}^-}$. (Proposition \ref{Prop:Density})
\end{itemize}

We will then combine these results in Theorem \ref{Thm:Fbijection} to obtain that $\mathcal{F}^-$ is a linear, bounded bijection with bounded inverse between the spaces  $\mathcal{E}^{X}_{\Sigma_{t^*_c}}$ and $\mathcal{E}^{\p_{t^*}}_{\mathcal{I}^-}$.

We will begin with the following:
\begin{Proposition}[Boundedness of $\mathcal{F}^-$]\label{Prop:BackScatBound}
	There exists a constant $A(\mathcal{M})$ such that
	\begin{equation}
	\Vert\mathcal{F}^-(\phi)\Vert_{\p_{t^*},\mathcal{I}^-}^2=\int_{\mathcal{I}^-}(\p_v(\mathcal{F}^-(\phi)))^2dvd\omega\leq A \Vert\phi\Vert_{\dot{H}^1(\Sigma_{t^*_c})}^2.
	\end{equation}
\end{Proposition}
\begin{proof}
	We will first prove this for compactly supported smooth functions, and then extend to $H^1$ functions using a density argument. 
	
	Let $X$, $w$ and $t^*_2$ be as in the proof of Theorem \ref{Thm:BackwardsBound}. Let $\phi$ be smooth and compactly supported on $\Sigma_{t^*_2}$. Take $v_0$ large enough such that on $\Sigma_{t^*_2}$, $\phi$ is only supported on $v\leq v_0$. We integrate $K^{X,w}$, in the region $D_{v_1}=\{v\in [v_1, v_0],t^*\leq t^*_2\}$, for any $v_1\leq v_0$.
	
	We then apply generalised Stokes' Theorem in $D_{v_1}$ to obtain the following boundary terms:
	\begin{align}\nonumber
	-\int_{\Sigma_{t^*_2}}dt^*(J^{X,w})&=\int_{D_{v_1}}K^{X,w}-\int_{\{v=v_1\}}dv(J^{X,w})+\int_{S_{[t^*_1,t^*_2]}}d\rho(J^{X,w})-\lim_{u_0\to-\infty}\int_{\{u=u_0\}\cap[v_1,v_0]}du(J^X,w)\\\label{eq:StokesonNull}
	&\geq-\int_{\{v=v_1\}}dv(J^{X,w})-\lim_{u_0\to-\infty}\int_{\{u=u_0\}\cap[v_1,v_0]}du(J^{X,w})
	\end{align}
	where $t^*_1$ is the value of $t^*$ at the sphere where $\{v=v_1\}$ intersects $\{r=r_b(t^*)\}$.
	
	\begin{align}\nonumber
	-\int_{\{v=v_1\}}dv(J^{X,w})=\int_{\{v=v_1\}}\frac{b}{2}&\left(\p_{t^*}\phi-\p_r\phi-\frac{\phi}{r}\right)^2\\\label{eq:dv(JXw)}
	&+\frac{1}{2}\left(\left(1-\frac{2M}{r}+\frac{q^2M^2}{r^2}\right)f(t^*)+\left(\frac{2M}{r}-\frac{q^2M^2}{r^2}\right)b\right)(\p_{t^*}\phi-\p_r\phi)^2\geq 0
	\end{align}
	\begin{align}\label{eq:duJXw}
	-\int_{\{u=u_0\}\cap[v_1,v_1]}du(J^{X,w})=&\int_{\{u=u_0\}\cap[v_1,v_0]}\frac{f(t^*)-b}{2\left(1-\frac{2M}{r}+\frac{q^2M^2}{r^2}\right)}\left(\left(1+\frac{2M}{r}-\frac{q^2M^2}{r^2}\right)\p_{t^*}\phi-\left(1-\frac{2M}{r}+\frac{q^2M^2}{r^2}\right)\p_r\phi\right)^2\\\nonumber
	&+\frac{b}{r}\phi\left(\frac{1+\frac{2M}{r}-\frac{q^2M^2}{r^2}}{1-\frac{2M}{r}+\frac{q^2M^2}{r^2}}\p_{t^*}\phi-\p_r\phi\right)-\frac{b\phi^2}{2r^2}+\frac{\left(1-\frac{2M}{r}+\frac{q^2M^2}{r^2}\right)f(t^*)-\left(1+\frac{2M}{r}-\frac{q^2M^2}{r^2}\right)b}{2r^2\left(1-\frac{2M}{r}+\frac{q^2M^2}{r^2}\right)}\vert\mathring{\slashed\nabla}\phi\vert^2
	\end{align}
	
	However, as we know that $r\phi$ tends to an $H^1_{loc}$ function, and the volume form on $\{u=u_0\}$ is $r^2$, we can see that the terms in \eqref{eq:duJXw} with a factor of $\phi$ tend to $0$ as $u_0\to\infty$. Similarly, by applying the rotational Killing fields $\Omega_i$ (defined in \eqref{eq:AngularKilling}) to $\phi$, we can see $r\Omega_i\phi$ has an $H^1_{loc}$ limit. Thus terms in \eqref{eq:duJXw} involving $\mathring{\slashed\nabla}\phi$ will also tend to $0$ in the limit $u_0\to\infty$. 
	
	Thus in the limit $u_0\to\infty$ (and therefore $r\to\infty$, $t^*\to-\infty$) we obtain:
	\begin{align}\nonumber
	-\lim_{u_0\to-\infty}\int_{\{u=u_0\}\cap[v_1,v_0]}du(J^{X,w})=&\lim_{u_0\to-\infty}\int_{\{u=u_0\}\cap[v_1,v_0]}\frac{1-b}{4}\left(1-\frac{2M}{r}+\frac{q^2M^2}{r^2}\right)\left(\p_{t^*}\phi-\p_r\phi\right)^2r^2dvd\omega\\\label{eq:duJXWI-}
	=&\lim_{u_0\to-\infty}\int_{\{u=u_0\}\cap[v_1,v_0]}\frac{1-b}{4}\left(1-\frac{2M}{r}+\frac{q^2M^2}{r^2}\right)\left(\p_{t^*}(r\phi)-\p_r(r\phi)\right)^2dvd\omega\\\nonumber=&\int_{\mathcal{I}^-\cap[v_1,v_0]}\frac{1-b}{4}\left(\p_{t^*}\psi_--\p_r\psi_-\right)^2dvd\omega
	\geq\epsilon\int_{\mathcal{I}-\cap[v_1,v_0]}\left(\p_v\psi_-\right)^2dvd\omega
	\end{align}
	where to get from the first line to the second, we have ignored terms of order $\phi$, as these tend to $0$.
	
	Substituting \eqref{eq:duJXWI-} and \eqref{eq:dv(JXw)} into \eqref{eq:StokesonNull}, and noting that $-\int_{\Sigma_{t^*_2}}dt^*(J^{X,w})$ can be bounded by the $\dot{H}^1$ norm, we have that:
	\begin{equation}
	\int_{\mathcal{I}^-\cap[v_0,v_1]}(\p_v(\mathcal{F}^-(\phi)))^2dvd\omega\leq A \Vert\phi\Vert_{\dot{H}^1(\Sigma_{t^*_2})}^2,
	\end{equation}
	where $A$ is independent of $v_1$. Thus taking a limit as $v_1\to-\infty$ and imposing Theorem $\ref{Thm:FiniteBound}$ in the region $[t^*_2,t^*_c]$ gives us the result of the theorem.
\end{proof}

We then move on to showing $\mathcal{F}^+$, if it exists, would be bounded:
\begin{Proposition}[Boundedness of $\mathcal{F}^+$]\label{Prop:ForwardScatBound}
		There exists a constant $A$ such that
	\begin{equation}
	\Vert\phi\Vert_{\dot{H}^1(\Sigma_{t^*_c})}^2\leq A \int_{\mathcal{I}^-}(\p_v(\mathcal{F}^-(\phi)))^2dvd\omega.
	\end{equation}
\end{Proposition}
To prove this, we will first need to show a decay result:
\begin{Lemma}[Decay of Solutions Along a Null Foliation]\label{Lem:Decay}
	Let $\phi$ be a solution to \eqref{eq:wave} with boundary conditions \eqref{eq:BoundaryConditions}. Then
	\begin{equation}
	\lim_{v_0\to-\infty}\int_{\Sigma_{v_0}}dv(J^{\p_{t^*}})=0.
	\end{equation}
\end{Lemma}
\begin{proof}
	We first show the result for $\phi$ compactly supported on some $\Sigma_{v_1}$, and then extend the result by a density argument.
	
	Firstly, we calculate $-dv(J^{\p_{t^*}})$ and $-du(J^{\p_{t^*}})$.
	\begin{align}\nonumber
	-\int_{\Sigma_{v_0}}dv(J^{\p_{t^*}})&=\frac{1}{2}\int_{\Sigma_{v_0}}\left(1-\frac{2M}{r}+\frac{q^2M^2}{r^2}\right)\left(\p_{t^*}\phi-d_r\phi\right)^2+\frac{1}{r^2}\vert\mathring{\slashed\nabla}\phi\vert^2\\
	&=\int_{\Sigma_{v_0}}2\left(1-\frac{2M}{r}+\frac{q^2M^2}{r^2}\right)^{-1}\left(\p_{u}\phi\right)^2+\frac{1}{2r^2}\vert\mathring{\slashed\nabla}\phi\vert^2=\int_{\Sigma_{v_0}}\left(\p_{u}\phi\right)^2+\frac{1-\frac{2M}{r}+\frac{q^2M^2}{r^2}}{4r^2}\vert\mathring{\slashed\nabla}\phi\vert^2r^2dud\omega
	\end{align}
	\begin{align}\label{eq:duJT}
	-\int_{\Sigma_{u_0}\cap[v_0,v_1]}du(J^{\p_{t^*}})&=\frac{1}{2}\int_{\Sigma_{u_0}\cap[v_0,v_1]}\left(1-\frac{2M}{r}+\frac{q^2M^2}{r^2}\right)\left(\left(\frac{1+\frac{2M}{r}-\frac{q^2M^2}{r^2}}{1-\frac{2M}{r}+\frac{q^2M^2}{r^2}}\right)\p_{t^*}\phi-\p_r\phi\right)^2+\frac{1}{r^2}\vert\mathring{\slashed\nabla}\phi\vert^2\\\nonumber
	&=\int_{\Sigma_{u_0}\cap[v_0,v_1]}\left(\p_{v}\phi\right)^2+\frac{1-\frac{2M}{r}+\frac{q^2M^2}{r^2}}{4r^2}\vert\mathring{\slashed\nabla}\phi\vert^2r^2dvd\omega\geq 0
	\end{align}
	
	Integrating $K^{\p_{t^*}}$ in the area $D_{u_0}:=\{v\in[v_0,v_1]\}\cap\{u\geq u_0\}$, using \eqref{eq:KT}, \eqref{eq:drho(JT)}, and then letting $u_0\to-\infty$ gives us that
	\begin{align}\label{eq:BoundedSigmav}
	-\int_{\Sigma_{v_1}}dv(J^{\p_{t^*}})\leq-\int_{\Sigma_{v_0}}dv(J^{\p_{t^*}}).
	\end{align}

	We then proceed to use the $r^p$ method \cite{NewPhysSpace}. We consider the wave operator applied to $r\phi$:
	\begin{align}\label{eq:rp}
	\frac{4\p_u\p_v(r\phi)}{1-\frac{2M}{r^2}+\frac{q^2M^2}{r^2}}=\frac{1}{r^2}\mathring{\slashed\triangle}(r\phi)-\frac{2M}{r^3}\left(1-\frac{q^2M}{r}\right)r\phi.
	\end{align}
	We apply \eqref{eq:rp} to the following integral over $D_{u_0}$:
	\begin{align}\nonumber
	\int_{\Sigma_{v_1}}\Bigg(\frac{2r}{1-\frac{2M}{r}+\frac{q^2M^2}{r^2}}&\left(\p_u(r\phi)\right)^2\Bigg)dud\omega\geq\left(\int_{\Sigma_{v_1}}-\int_{\Sigma_{v_0}}-\int_{S_{[v_0,v_1]}}\right)\left(\frac{2r}{1-\frac{2M}{r}+\frac{q^2M^2}{r^2}}\left(\p_u(r\phi)\right)^2\right)dud\omega\\\nonumber
	=&\int_{D_{u_0}}\frac{4r\p_u(r\phi)\p_u\p_v(r\phi)}{1-\frac{2M}{r^2}+\frac{q^2M^2}{r^2}}+2\left(\p_u(r\phi)\right)^2\p_v\left(\frac{1}{r}-\frac{2M}{r^2}+\frac{q^2M^2}{r^2}\right)^{-1}dudvd\omega\\\label{eq:rpmethod}
	=&\int_{D_{u_0}}\left(\frac{1}{r}\mathring{\slashed\triangle}(r\phi)-\frac{2M}{r^2}\left(1-\frac{q^2M}{r}\right)r\phi\right)\p_u(r\phi)+\frac{\left(1-\frac{4M}{r}+\frac{3q^2M^2}{r^2}\right)(\p_u(r\phi))^2}{1-\frac{2M}{r}+\frac{q^2M^2}{r^2}}dudvd\omega\\\nonumber
	=&\int_{D_{u_0}}\frac{\left(1-\frac{4M}{r}+\frac{3q^2M^2}{r^2}\right)(\p_u(r\phi))^2}{1-\frac{2M}{r}+\frac{q^2M^2}{r^2}}-\frac{1}{2r}\p_u\left(\vert\mathring{\slashed\nabla}r\phi\vert^2\right)-\frac{M}{r^2}\left(1-\frac{q^2M}{r}\right)\p_u((r\phi)^2)dudvd\omega\\\nonumber
	\geq&\int_{D_{u_0}}\frac{\left(1-\frac{4M}{r}+\frac{3q^2M^2}{r^2}\right)(\p_u(r\phi))^2}{1-\frac{2M}{r}+\frac{q^2M^2}{r^2}}+\left(1-\frac{2M}{r}+\frac{q^2M^2}{r^2}\right)\left(\frac{1}{4r^2}\vert\mathring{\slashed\nabla}r\phi\vert^2+\left(1-\frac{3q^2M}{2r}\right)\frac{M}{r^3}(r\phi)^2\right)dudvd\omega\\\nonumber
	\geq&\frac{1}{2}\int_{D_{u_0}}(\p_u\phi)^2+\frac{1-\frac{2M}{r}+\frac{q^2M^2}{r^2}}{2r^2}\vert\mathring{\slashed\nabla}\phi\vert^2r^2dudvd\omega\geq\frac{1}{2}\int_{v_0}^{v_1}\left(-\int_{\Sigma_{v_0}}dv(J^{\p_{t^*}})\right)dv
	\end{align}
	In order to obtain the last line, we have used that for $r$ large enough,
	\begin{align}\nonumber
	\int_{\Sigma_{v}}\frac{1-\frac{4M}{r}+\frac{3q^2M^2}{r^2}}{1-\frac{2M}{r}+\frac{q^2M^2}{r^2}}(\p_u(r\phi))^2dud\omega&\geq\frac{1}{2}\int_{\Sigma_{v}}(\p_u(r\phi))^2dud\omega=\frac{1}{2}\int_{\Sigma_{v}}r^2(\p_u\phi)^2+\p_u\left(r\p_u r \phi^2\right)-\p_u^2 r\frac{(r\phi)^2}{r}dud\omega\\
	&\geq\frac{1}{2}\int_{\Sigma_{v}}r^2(\p_u\phi)^2-\frac{1}{2}\left(1-\frac{2M}{r}+\frac{q^2M^2}{r^2}\right)\left(1-\frac{q^2M}{r}\right)\frac{M}{r^2}\phi^2dud\omega\\\nonumber
	&\geq\frac{1}{2}\int_{\Sigma_{v}}r^2(\p_u\phi)^2-\frac{M}{2r^2}\phi^2dud\omega.
\end{align}

The left hand side of \eqref{eq:rpmethod} is independent of $v_0$, so if we choose $\phi$ to be compactly supported on $\Sigma_{v_1}$ (these functions are dense in the set of $H^1$ functions), then we can let $v_0$ tend to $-\infty$ to obtain
\begin{equation}
\int_{-\infty}^{v_1}\left(	-\int_{\Sigma_{v_0}}dv(J^{\p_{t^*}})\right)dv\leq\int_{\Sigma_{v_1}}\left(\frac{4r}{1-\frac{2M}{r}+\frac{q^2M^2}{r^2}}\left(\p_u(r\phi)\right)^2\right)dud\omega.
\end{equation}

Thus there exists a sequence $v_i\to-\infty$ such that
\begin{equation}\label{eq:DecayingSequence}
	-\int_{\Sigma_{v_i}}dv(J^{\p_{t^*}})\to 0\quad \text{ as } i\to\infty.
\end{equation}

We then note that given any $\epsilon>0$, and a solution $\phi$ to \eqref{eq:wave} with finite $\p_{t^*}$ energy on $\Sigma_{v_0}$, there exists a smooth compactly supported function $\phi_\epsilon$ such that
\begin{equation}
\Vert\phi-\phi_{\epsilon}\Vert_{\p_{t^*},\Sigma_{v_0}}\leq\epsilon/2.
\end{equation}

Furthermore, by \eqref{eq:BoundedSigmav}, we know that for all $v_1\leq v_0$, we have
\begin{equation}\label{eq:DecayApprox}
\Vert\phi-\phi_{\epsilon}\Vert_{\p_{t^*},\Sigma_{v_1}}\leq\epsilon/2.
\end{equation}

By \eqref{eq:DecayingSequence}, there exists a $v_1\leq v_0$ such that
\begin{equation}\label{eq:DecaySmooth}
\Vert \phi_\epsilon\Vert_{\p_{t^*},\Sigma_{v_1}}\leq\epsilon/2.
\end{equation}

By combining \eqref{eq:DecayApprox} \eqref{eq:DecaySmooth}, and \eqref{eq:BoundedSigmav} again, we obtain that
\begin{equation}
\Vert\phi\Vert_{\p_{t^*},\Sigma_{v}}\leq\Vert\phi\Vert_{\p_{t^*},\Sigma_{v_1}}\leq\Vert\phi_\epsilon\Vert_{\p_{t^*},\Sigma_{v_1}}+\Vert\phi-\phi_\epsilon\Vert_{\p_{t^*},\Sigma_{v_1}}\leq\epsilon,
\end{equation}
for all $v\leq v_1$.

Thus given any solution $\phi$ to \eqref{eq:wave} with finite $\p_{t^*}$ energy, and given any $\epsilon>0$, there exists a $v_1$ such that
\begin{equation}
\Vert\phi\Vert_{\p_{t^*},\Sigma_{v}}\leq\epsilon,
\end{equation}
for all $v\leq v_1$.
\end{proof}
\begin{proof}[Proof of Proposition \ref{Prop:ForwardScatBound}]
	Fix $t^*_1<t^*_c$, and let $\phi$ be a solution of \eqref{eq:wave} with boundary conditions \eqref{eq:BoundaryConditions} such that $\phi$ has finite $\p_{t^*}$ energy on $\Sigma_{t^*_1}$. Note as $t^*_1<t^*_c$ fixed, finite $\p_{t^*}$ energy is equivalent to having finite $X$ energy. An explicit calculation gives
	\begin{equation}
	\left(1-\frac{2M}{r_b(t^*_1)}+\frac{q^2M^2}{r_b(t^*_1)^2}\right)\Vert\phi\Vert^2_{\dot{H}^1(\Sigma_{t^*_1})}\leq-\int_{\Sigma_{t^*_1}}dt^*(J^{\p_{t^*}})\leq\Vert\phi\Vert^2_{\dot{H}^1(\Sigma_{t^*_1})}
	\end{equation} 
	
	We prove Proposition \ref{Prop:ForwardScatBound} by simply integrating $K^{\p_{t^*}}$ in the region $D_{u_0,v_0}=\{u>u_0, v\geq v_0,t^*<t^*_1\}$. We will then let $u_0\to-\infty$ to get:
	\begin{align}
	-\int_{\Sigma_{t^*_1}}dt^*(J^{\p_{t^*}})=&\lim_{u_0\to-\infty}\left(-\int_{\Sigma_{t^*_1}\cap\{u\geq u_0\}}dt^*(J^{\p_{t^*}})\right)\\\nonumber
	=&-\lim_{u_0\to-\infty}\int_{\Sigma_{u_0}\cap\{v\geq v_0\}}du(J^{\p_{t^*}})-\lim_{u_0\to-\infty}\int_{\Sigma_{v_0}\cap\{u\geq u_0\}}dv(J^{\p_{t^*}})-\int_{S[v_0,t^*_1]}d\rho(J^{\p_{t^*}})\\\nonumber
	\leq&\int_{\mathcal{I}^-\cap\{v\geq v_0\}}(\p_v\psi_-)^2dvd\omega-\int_{\Sigma_{v_0}}dv(J^{\p_{t^*}}).
	\end{align}
	Here we have used \eqref{eq:drho(JT)} to ignore the $S[v_0,t^*_1]$ term. Letting $v_0\to-\infty$, and using Lemma \ref{Lem:Decay}, we obtain
	\begin{equation}
		\left(1-\frac{2M}{r_b(t^*_1)}+\frac{q^2M^2}{r_b(t^*_1)^2}\right)\Vert\phi\Vert^2_{\dot{H}^1(\Sigma_{t^*_1})}\leq-\int_{\Sigma_{t^*_1}}dt^*(J^{\p_{t^*}})\leq\int_{\mathcal{I}^-}(\p_v\psi_-)^2dvd\omega.
	\end{equation} 
	
	Theorem \ref{Thm:FiniteBound} on the interval $t^*\in[t^*_1,t^*_c]$ then gives us our result.
\end{proof}

\begin{wrapfigure}{r}{5cm}
	\begin{tikzpicture}[scale =1.2]
	\node (I)    at ( 0,0) {};
	\path 
	(I) +(0:2)   coordinate[label=0:$i^0$] (Iright)
	+(-1,1) coordinate (Ileft)
	+(-0.4,1) coordinate (BHH)
	+(-1,-3) coordinate[label=0:$i^-$] (Ibot)
	;
	\draw (Iright) -- 
	node[midway, below, sloped] {$\psi_-$}
	(Ibot) --
	node[midway, above, sloped]    {\small }    
	(Ileft);
	\draw[fill=gray!80] (Ibot) to[out=70, in=-90]
	node[midway, below, sloped] {\tiny $r=r_b$} (BHH)--(Ileft)--cycle;
	\draw (0,-2) to node[midway, below, sloped] {\tiny $\Sigma_{v_0}$} (-0.56,-1.44);
	\draw (0,-2) to[out=45,in=-135] (0.9,-0.9) to[out=45,in=-135] (1.5,-0.5);
	\draw (-0.6,-1.6) to[out=20,in=-150] node[midway, above, sloped] {\tiny $\Sigma_{t^*_0}$, $\phi'\vert_{\Sigma_{t^*_0}}$} (Iright);
	\end{tikzpicture}
\end{wrapfigure}

We now have that $\mathcal{F}^-$ is bounded and injective, so the inverse is well defined. We also have that $\mathcal{F}^+$ is bounded where it is defined. The final result needed to define the scattering map on the whole space $\mathcal{E}^{\p_{t^*}}_{\mathcal{I}^-}$ is that the image of $\mathcal{F}^-$ is dense in $\mathcal{E}^{\p_{t^*}}_{\mathcal{I}^-}$:

\begin{Proposition}[Density of $Im(\mathcal{F}^-)$ in $\mathcal{E}^{\p_{t^*}}_{\mathcal{I}^-}$]\label{Prop:Density}
	$Im(\mathcal{F}^-)$ is dense in $\mathcal{E}^{\p_{t^*}}_{\mathcal{I}^-}$.
\end{Proposition}
\begin{proof}
	We prove this using existing results on the scattering map on the full exterior of Reissner--Nordstr\"om spacetime. We show that compactly supported smooth functions on $\mathcal{I}^-$ are in the image of $\mathcal{F}^-$. These are dense in $\mathcal{E}^{\p_{t^*}}_{\mathcal{I}^-}$.
	
	Given any smooth compactly supported function $\psi_-\in\mathcal{E}^{\p_{t^*}}_{\mathcal{I}^-}$, supported in $v\geq v_0$, we can find a $t^*_0$ such that the sphere $(t^*_0,r_b(t^*_0))$ is in the region $v\leq v_0$. Using previous results from \cite{HafnerD2001Acft}, there exists a solution, $\phi'$ in Reissner--Nordstr\"om with radiation field $\psi$ and vanishing on the past horizon. By finite speed of propagation, $\phi'$ will be supported in $v\geq v_0$. Thus both $\phi'$ and its derivatives on $\Sigma_{t^*_0}$ vanishes around $r=r_b$.
	
	We then evolve $(\phi'\vert_{\Sigma_{t^*_0}},\p_{t^*}\phi'\vert_{\Sigma_{t^*_0}})$ from $\Sigma_{t^*_0}$ in RNOS, call this solution $\phi$. By finite speed of propagation and uniqueness of solutions, we must have $\phi=\phi'$ for $t^*\leq t^*_0$. By boundedness of $\mathcal{F}^+$ (Proposition \ref{Prop:ForwardScatBound}) we have that $\phi$ is in $\mathcal{E}^X_{\Sigma_{t^*_c}}$, and so the radiation field, $\psi_-$, is in the image of $\mathcal{F}^-$.
\end{proof}

\begin{Theorem}[Bijectivity and Boundedness of $\mathcal{F}^-$]\label{Thm:Fbijection}
	$\mathcal{F}^-$ is a linear, bounded bijection with bounded inverse between the spaces $\mathcal{E}^X_{\Sigma_{t^*_c}}$ and $\mathcal{E}^{\p_{t^*}}_{\mathcal{I}^-}$.
\end{Theorem}
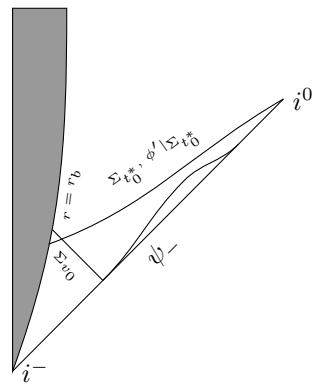
\begin{proof}
	$\mathcal{F}^+$ is continuous (linear and bounded, by Proposition \ref{Prop:ForwardScatBound}), and its image, $\mathcal{E}^X_{\Sigma_{t^*_c}}$, is a closed set. Therefore $\left(\mathcal{F}^+\right)^{-1}\left(\mathcal{E}^X_{\Sigma_{t^*_c}}\right)$ is closed. Thus
	\begin{equation}
	\mathcal{F}^-(\mathcal{E}^X_{\Sigma_{t^*_c}})=\left(\mathcal{F}^+\right)^{-1}\left(\mathcal{E}^X_{\Sigma_{t^*_c}}\right)=Cl\left(\left(\mathcal{F}^+\right)^{-1}\left(\mathcal{E}^X_{\Sigma_{t^*_c}}\right)\right)\supset\mathcal{E}^{\p_{t^*}}_{\mathcal{I}^-},
	\end{equation}
	as $\mathcal{F}^-(\mathcal{E}^X_{\Sigma_{t^*_c}})$ is dense in $\mathcal{E}^{\p_{t^*}}_{\mathcal{I}^-}$ (Proposition \ref{Prop:Density}).
	
	However, as $\mathcal{F}^-$ is also bounded, we have 
	\begin{equation}
	\mathcal{F}^-(\mathcal{E}^X_{\Sigma_{t^*_c}})\subset\mathcal{E}^{\p_{t^*}}_{\mathcal{I}^-}
	\end{equation}
	
	Therefore 
	\begin{equation}
	\mathcal{F}^-(\mathcal{E}^X_{\Sigma_{t^*_c}})=\mathcal{E}^{\p_{t^*}}_{\mathcal{I}^-}.
	\end{equation}
	
	Thanks to Propositions \ref{Prop:BackScatBound} and \ref{Prop:ForwardScatBound}, $\mathcal{F}^-$ and $\mathcal{F}^+$ are bounded, and thus we have the required result.
\end{proof}

\subsection{Forward Scattering from $\Sigma_{t^*_c}$}
In a similar manner to Section \ref{Sec:Backwards Scattering}, we define the map taking initial data on $\Sigma_{t^*_c}$ to radiation fields on $\mathcal{H}^+\cup\mathcal{I}^+$:
\begin{align}
\mathcal{G}^+:\ &\ \ \ \ \ \  \mathcal{E}^X_{\Sigma_{t^*_c}}\ \ \ \ \ \ \ \  \longrightarrow \ \ \ \mathcal{G}^+\left(\mathcal{E}^X_{\Sigma_{t^*_c}}\right)\subset H^1_{loc}(\mathcal{H}^+\cup\mathcal{I}^+)\\\nonumber
&(\phi|_{\Sigma_{t^*_c}},\p_{t^*}\phi|_{\Sigma_{t^*_c}})\mapsto \ \ \ (\phi|_{\mathcal{H}^+}, \psi_+)
\end{align}
where $\psi_+$ is as in Proposition \ref{Prop:RadFieldExist}, and $X$ is again any everywhere timelike vector field (including on the event horizon) which coincides with the timelike Killing vector field $\p_{t^*}$ for sufficiently large $r$. We will define the inverse of $\mathcal{G}^+$ (only defined on the image of $\mathcal{G}^+$) as 
\begin{align}
	\mathcal{G}^-:&\mathcal{G}^+\left(\mathcal{E}^X_{\Sigma_{t^*_c}}\right)\qquad\longrightarrow\qquad \mathcal{E}^X_{\Sigma_{t^*_c}}\\\nonumber
	\mathcal{G}^+&(\phi|_{\Sigma_{t^*_c}},\p_{t^*}\phi|_{\Sigma_{t^*_c}})\mapsto(\phi|_{\Sigma_{t^*_c}},\p_{t^*}\phi|_{\Sigma_{t^*_c}}).
\end{align}

\begin{Remark}
Note that $\mathcal{G}^\pm$ are defined using scattering in pure Reissner--Nordstr\"om. Thus they have been studied extensively already, see for example \cite{BHL} for the sub-extremal case ($\vert q\vert<1$) and \cite{ERNScat} for the extremal case ($\vert q\vert=1$).
\end{Remark}

We will be using the following facts about $\mathcal{G}^+$:
\begin{Lemma}\label{Lem:G+Facts}
	\begin{itemize}
		\item $\mathcal{G}^+$ is injective.
		\item For the sub-extremal case ($\vert q\vert<1$), $\mathcal{G}^+$ is bounded with respect to the $X$ norm on $\Sigma_{t^*_c}$ and $\mathcal{H}^+$ and the $\p_{t^*}$ norm on $\mathcal{I}^+$. In the extremal case ($\vert q\vert=1$), we use the weaker result that $\mathcal{G}^+$ is bounded with respect to the $X$ norm on $\Sigma_{t^*_c}$ and the $\p_{t^*}$ norm on $\mathcal{I}^+$ and $\mathcal{H}^+$.
		\item $\mathcal{G}^+$ is not surjective into $\mathcal{E}^{\p_{t^*}}_{\mathcal{I}^+}$, for both sub-extremal and extremal Reissner--Nordstr\"om. %$Im(\mathcal{G}^+)$ does not even include $0\times \mathcal{E}^{\p_{t^*}}_{\mathcal{I}^+}$.
		\item $\mathcal{G}^-$ is not bounded, again with respect to the $X$ norm on $\Sigma_{t^*_c}$ and $\mathcal{H}^+$, and the $\p_{t^*}$ norm on $\mathcal{I}^+$.
	\end{itemize}
\end{Lemma}
\begin{proof}
	Thanks to $T$ energy conservation, we have that $\mathcal{G}^+$ is injective. 
	
	For $\mathcal{G}^+$ bounded in the sub-extremal case, we apply the celebrated red-shift vector \cite{RedShift} in order to obtain boundedness of the $X$ energy on $\mathcal{H}^+$. 
	
	In the extremal case, we do not have the red-shift effect. In this case, the best we can do is apply conservation of $T$ energy which immediately gives the weaker extremal result.
	
	For $\mathcal{G}^+$ not surjective, we can look at any solution with finite $\p_{t^*}$ energy on $\Sigma_{t^*_0}$, but infinite $X$ energy, such as $\phi=\sqrt{r-r_+}$, $\p_{t^*}\phi=0$. Let $\mathcal{G}^+(\phi)= (\phi_+,\psi_+)$, which has finite $X$ and $\p_{t^*}$ energy respectively (the angular component vanishes by spherical symmetry). $\mathcal{G}^+$ is injective from the $\p_{t^*}$ energy space, thus no other finite $\p_{t^*}$ energy data on $\Sigma_{t^*_c}$ can map to $(\phi_+,\psi_+)$. Therefore no finite $X$ energy solution can map to $(\phi_+,\psi_+)$, and thus $\mathcal{G}^+\left(\mathcal{E}^X_{\Sigma_{t^*_c}}\right)\neq\mathcal{E}^X_{\mathcal{H}^+}\times\mathcal{E}^{\p_{t^*}}_{\mathcal{I}^+}$. For a more detailed discussion of non-surjectivity in the sub-extremal case see \cite{Blue} (note this proves non-surjectivity for Kerr, but the proof can be immediately applied to Reissner--Nordstr\"om). For the extremal case, again see \cite{ERNScat}.
	
	By taking a series of smooth compactly supported functions approximating $(\phi_+,\psi_+)$ in the above paragraph, we can see that $\mathcal{G}^-$ is not bounded.
\end{proof}

\subsection{The Scattering Map}
We are finally able to define the forwards Scattering Map:
\begin{align}\label{eq:ForwardScatteringMap}
\mathcal{S}^+:\mathcal{E}^{\p_{t^*}}_{\mathcal{I}^-}&\longrightarrow \mathcal{S}^+\left(\mathcal{E}^{\p_{t^*}}_{\mathcal{I}^-}\right)\\\nonumber
\mathcal{S}^+&:=\mathcal{G}^+\circ\mathcal{F}^+
\end{align}
and similarly with the backwards scattering map:
\begin{align}
\mathcal{S}^-:\mathcal{S}^+\Big(&\mathcal{E}^{\p_{t^*}}_{\mathcal{I}^-}\Big)\longrightarrow \mathcal{E}^{\p_{t^*}}_{\mathcal{I}^-}\\\nonumber
\mathcal{S}^-&:=\mathcal{F}^-\circ\mathcal{G}^-.
\end{align}
Note $\mathcal{S}^-$ is defined only on the image of $\mathcal{S}^+$.
\begin{Theorem}[The Scattering Map]\label{Thm:ScatBound}
	The sub-extremal ($\vert q\vert<1$) forward scattering map $\mathcal{S}^+$ defined by \eqref{eq:ForwardScatteringMap} is an injective linear bounded map from $\mathcal{E}^{\p_{t^*}}_{\mathcal{I}^-}$ to $\mathcal{E}^X_{\mathcal{H}^+}\cup\mathcal{E}^{\p_{t^*}}_{\mathcal{I}^+}$. The extremal ($\vert q\vert=1$) forward scattering map $\mathcal{S}^+$, again defined by \eqref{eq:ForwardScatteringMap}, is an injective linear bounded map from $\mathcal{E}^{\p_{t^*}}_{\mathcal{I}^-}$ to $\mathcal{E}^{\p_{t^*}}_{\mathcal{H}^+}\cup\mathcal{E}^{\p_{t^*}}_{\mathcal{I}^+}$. In both cases, $\mathcal{S}^+$ is not surjective, and its image does not even contain ${0}+\mathcal{E}^{\p_{t^*}}_{\mathcal{I}^+}$. When defined on the image of $\mathcal{S}^+$, its inverse $\mathcal{S}^-$ is injective but not bounded.
\end{Theorem}
\begin{proof}
This is an easy consequence of Theorem \ref{Thm:Fbijection} and Lemma \ref{Lem:G+Facts}.
\end{proof}

This is in immediate contrast with Reissner--Nordstr\"om spacetime. The scattering map in Reissner--Nordstr\"om spacetime is an isometry with respect to the $T$ energy, and this immediately follows from the fact that $T$ is a global Killing vector field. Moreover, this imposes the canonical choice of energy on $\mathcal{I}^\pm$. 

However, in the RNOS model, if one considers the $T$ energy on $\mathcal{I}^-$, then $\mathcal{F}^+$ gives an isometry between $\mathcal{E}_{\mathcal{I}^-}^{T}$ and $\mathcal{E}_{\Sigma_{t^*_c}}^{X}$. Thus, we are forced to consider the non-degenerate $X$ energy, when considering the solution on $\Sigma_{t^*_c}$. This is the main contrast with Reissner--Nordstr\"om spacetime, where we consider $T$ energy throughout the whole spacetime.

In both Reissner--Nordstr\"om and RNOS spacetimes, we can consider the backwards reflection map, which takes finite energy solutions from $\mathcal{I}^+$ to $\mathcal{I}^-$. On both these surfaces, choice of energy is canonically given by the existence of Killing vector fields in the region around $\mathcal{I}^\pm$. In Reissner--Nordstr\"om this map is bounded, however in RNOS, this map does not even exist as a map between finite energy spaces.

\bibliographystyle{unsrt}
\bibliography{SES-O.bib}

\end{document}